\documentclass{ar-1col}

\usepackage{amsmath}
\usepackage{amsfonts}
\usepackage{amssymb}
\usepackage{graphicx}
\usepackage{algorithm} 
\usepackage{algpseudocode}
\usepackage{natbib}
\usepackage[margin=1.3in]{geometry}
\usepackage{bm}
\usepackage{bbm}
\usepackage{hyperref}
\usepackage{placeins}
\usepackage{amsthm}
\usepackage{caption,subcaption}
\usepackage{ulem}
\usepackage{epstopdf}
\usepackage{wrapfig,lipsum}

\jname{Annu. Rev. Stat. Appl.}
\jvol{5}
\jyear{2018}
\doi{10.1146/((please add article doi))}

\begin{document}

\markboth{Barp, Briol, Kennedy, and Girolami}{Geometry \& Dynamics for Markov Chain Monte Carlo}

\title{Geometry \& Dynamics for Markov Chain Monte Carlo}

\author{Alessandro Barp$^{1,*}$, Fran\c{c}ois-Xavier Briol$^{2,*}$, Anthony D. Kennedy$^{3,*}$ and Mark Girolami$^{4,*}$
\affil{$^1$Department of Mathematics, Imperial College London, London, United Kingdom, SW7 2AZ; email: a.barp16@imperial.ac.uk}
\affil{$^2$Department of Statistics, University of Warwick, Coventry, United Kingdom, CV4 7AL; email: f-x.briol@warwick.ac.uk}
\affil{$^3$School of Physics and Astronomy, University of Edinburgh, Edinburgh, United Kingdom, EH9 3JZ; email: adk@ph.ed.ac.uk}
\affil{$^4$Department of Mathematics, Imperial College London, London, United Kingdom, SW7 2AZ; email: m.girolami@imperial.ac.uk}
\affil{$^*$The Alan Turing Institute, British Library, London, United Kingdom, NW1 2DB.}
}

\begin{abstract}

Markov Chain Monte Carlo methods have revolutionised mathematical computation and enabled statistical inference within many previously intractable models. In this context, Hamiltonian dynamics have been proposed as an efficient way of building chains which can explore probability densities efficiently. 
The method emerges from  physics and geometry and these links have been extensively studied by a series of authors through the last thirty years. However, there is currently a gap between the intuitions and knowledge of users of the methodology and our deep understanding of these theoretical foundations. The aim of this review is to provide a comprehensive introduction to the geometric tools used in Hamiltonian Monte Carlo at a level accessible to statisticians, machine learners and other users of the methodology with only a basic understanding of Monte Carlo methods. This will be complemented with some discussion of the most recent advances in the field which we believe will become increasingly relevant to applied scientists.
\end{abstract}

\maketitle

\setlength{\parskip}{3pt}%

\section{INTRODUCTION}

\subsection{Markov Chain Monte Carlo}

One of the aims of Monte Carlo methods is to sample from a target distribution, that is, to generate a set of identically independently distributed (i.i.d) samples $x^{(i)}$ with respect to the density $\pi$ of this distribution. 
Sampling from such a distribution enables the estimation of the integral $\mathbb{E}_{\pi}[f]=\int_{\mathcal{X}} fd\Pi $ of a function $f:\mathcal{X}\rightarrow \mathbb{R}$ with respect to its corresponding probability measure $\Pi$ by $\frac{1}{m}\sum_i^mf\big{(}x^{(i)}\big{)}$. Formally, the target density is a non-negative (almost everywhere) measurable function $\pi:\mathcal{X}\rightarrow \mathbb{R}^+$, where $\mathcal{X}\subset \mathbb{R}^d$ is the sample space of a measurable space with Lebesgue measure $\mu$, corresponding to the probability measure $\Pi(A)=\int_A \pi d\mu$.

Often we only know $\pi$ up to a multiplicative constant, that is we are able to evaluate $\tilde{\pi}$ where $\pi=\tilde{\pi}/Z$ for some $Z \in \mathbb{R}^+$. For example, this is the case in Bayesian statistics, where the normalisation constant $Z$ is the model evidence, which is itself a complicated integral not always available in closed form. Even when we know the value of $Z$, sampling from $\pi$ is challenging, particularly in high dimensions where high probability regions are usually concentrated on small subsets of the sample space \citep{MacKay2003}. There are only few densities for which we can easily generate samples.

The first Markov Chain Monte Carlo (MCMC) algorithm appeared in physics \cite{Metropolis1953} as a way of tackling these issues. The problem investigated was a large system of particles, and the aim was to compute the expected value of physical quantities. The high dimension of the system made it impossible to use numerical methods or standard Monte Carlo to compute the integral. Instead they proposed a method based on generating samples from an arbitrary random walk, and adding an accept/reject step to ensure they originate from the correct distribution.
Despite extensive use in statistical mechanics, the first mention of MCMC in the statistical literature appeared twenty years later \citep{Hasting1970}. It is a paper from \cite{Gelfand1990} that finally set things moving and marks the beginning of the MCMC revolution in statistics \citep{Robert2011}.

The idea behind MCMC methods \citep{Meyn1993,Robert2004} is to generate samples from the target $\pi$ which are approximately i.i.d. by defining a Markov chain whose stationary density is $\pi$. Recall that a Markov chain is a sequence of random variables $(X_0,X_1,\ldots)$ such that the distribution of $X_r$ depends only on $X_{r-1}$. A Markov chain may be specified by an \textbf{initial density} $h_0(x)$ for $X_0$ and a \textbf{transition density} $T(x' \leftarrow x)$ from which we can sample. The density of $X_r$ is then defined by $h_r(x')=\int T(x' \leftarrow x)h_{r-1}(x)dx$. 
The density $\pi$ is called a \textbf{stationary density} of the Markov chain if whenever $X_r \sim \pi$ then $X_{r+1} \sim \pi$, that is $\pi (x')=\int T(x' \leftarrow x)\pi(x)dx$. If the Markov chain is ergodic, it will converge to its stationary distribution independently of its initial distribution. A common way to guarantee $\pi$ is indeed the invariant density of the chain (which then asymptotically generates samples from $\pi$), is to demand that it satisfies the detailed balance condition $\pi(x)T(x'\leftarrow x) \; = \; \pi(x')T(x \leftarrow x')$. Intuitively, this condition requires that the probabilities of moving from state $x$ to $x'$ and from $x'$ to $x$ are equal. Note however that detailed balance is a sufficient but not necessary condition \citep{Diaconis2000}. 

The \textbf{Metropolis-Hastings algorithm} constructs a Markov chain converging to the desired target $\pi$ by the means of a proposal kernel $P$, where for each $x \in \mathcal{X}$, $P(\cdot,x)$ is a density on the state space from which we can sample. Given the current state $x_r$: 
\begin{enumerate}
\item Propose a new state $y \sim P(\cdot,x_r)$. 
\vspace{1mm}
\item Accept $y$ with probability $A(y|x_r) := \min \left\{1,\frac{\pi(y)P(x_r,y)}{\pi(x_r)P(y,x_r)}\right\}$, else set $x_{r+1}=x_r$.
\end{enumerate}
This induces a transition density $T(y \leftarrow x):=P(y,x) A(y|x_r)+\bm{1}_{\{y=x_r\}}(1-A(y|x_r))$, where $\bm{1}_{\{y=x_r\}}:\mathcal{X}\rightarrow \{0,1\}$ takes value $1$ when $y=x_r$ and $0$ otherwise. We emphasise that this quantity does not rely on the normalisation constant $Z$ since it cancels out in the ratio.

\subsection{Motivation for the Use of Geometry}

In principle, there are only mild requirements on the proposal $P$ to obtain an asymptotically correct algorithm, however this choice will be very significant for the performance of the algorithm. Intuitively, the aim is to choose a proposal which will favour values with high probability of acceptance whilst also exploring the state space well (i.e., have small correlations with the current state). A common choice is a symmetric density (e.g., Gaussian) centred on the current state of the chain, leading to an algorithm named \textbf{random-walk Metropolis} (RWM). A more advanced algorithm is the \textbf{Metropolis-Adjusted Langevin Algorithm} (MALA) \citep{Rossky1978,Scalettar1986,Roberts1998}, which uses the path of a diffusion which is invariant to the target distribution. 

As previously discussed, concentration of measure is a well-known phenomena in high dimensions \citep{Ledoux2001} and is linked to concentration of volume (also commonly referred to as the curse of dimensionality). An intuitive example, often used to describe this phenomenon, is that of a sphere $S^d$ embedded in the unit cube. One can show that most of the volume of the cube is concentrated outside the sphere, and this is increasingly the case for higher $d$. Similarly, probability measures will tend to concentrate around their mean in high dimensions \citep{MacKay2003,Betancourt2017}, making the use of RWM inefficient since it does not adapt to the target distribution.

\begin{figure}
\begin{subfigure}[b]{0.6\textwidth}
\includegraphics[width=0.8\textwidth,clip,trim = 0 0 0 0]{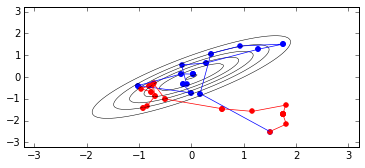}
\caption{}
\end{subfigure}
\begin{subfigure}[b]{0.4\textwidth}
\begin{center}
\includegraphics[width=0.7\textwidth,clip,trim = 0 0 0 0]{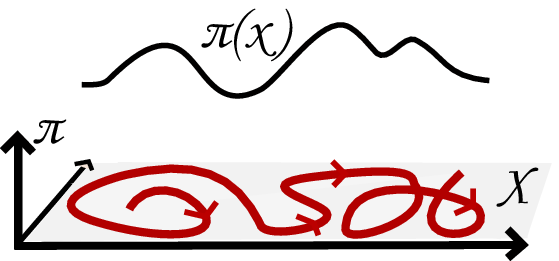}
\hspace{5mm}
\end{center}
\caption{}
\end{subfigure}
\hspace{0.5mm}
\caption{\textbf{(a)} Comparison of RWM (red) against HMC (blue). Thirty samples from a peaked Gaussian distribution were plotted for each method. The use of geometry clearly benefits HMC. \textbf{(b)} Motion of a particle (in red) over our sample space $\mathcal X$.}
\label{fig:simulation_HMC_RWM}
\end{figure}

To avoid issues with high curvature and concentration of measure, \citep{Duane1987} proposed a method based on approximately simulating from some Hamiltonian dynamics with potential energy given by the log-target density. Informally, this has the advantage of directing the Markov chain towards areas of high probability and hence providing more efficient proposals (see Fig.\ref{fig:simulation_HMC_RWM}-a.). This method was originally named Hybrid Monte Carlo (HMC), as it was an "hybrid" of Molecular Dynamics (microcanonical) and momentum heatbath (Gibbs sampler). The method is now also commonly known as Hamiltonian Monte Carlo \citep{Neal2011}.

HMC has been used throughout statistics but has also spanned a wide range of fields including Biology \citep{Berne1997,Hansmann1999,Kramer2014}, Medicine \citep{Konukoglu2011,Schroeder2013}, Computer Vision \citep{Choo2001}, Chemistry \citep{Ajay1998,Fredrickson2002,Fernandez-Pendas2014}, Physics \citep{Duane1987,Mehlig1992,Landau2009,Sen2017} and Engineering \citep{Cheung2009,Bui-Thanh2014,Lan2016,Beskos2016}. The extent of the use of HMC is also illustrated by the long list of users of the STAN language \citep{Carpenter2016}; see for example \url{http://mc-stan.org/citations} for a full list of publications referencing this software. The above is of course a far-from-exhaustive list, but it helps illustrate the relevance of HMC in modern computational sciences.

\subsection{Outline}

The remainder of this paper reviews the use of Hamiltonian dynamics in the context of MCMC. Previous reviews of this methodology were provided by \cite{Neal2011,Betancourt2017}, however they focussed mainly on the intuition and algorithmic aspects behind the basic version of HMC. Our aim here is somewhat different and complementary: we will focus on formalising the geometrical and physical foundations of the method (see \S \ref{sec:geometry_physics} \& \ref{sec:Hamiltonian_Monte_Carlo}). This deeper theoretical understanding has provided insight into the development of many extensions of HMC \citep{Betancourt2014GeomFoundations}. These include Riemannian Manifold Hamiltonian Monte Carlo (RMHMC) \citep{Girolami2011}, introduced in \S \ref{sec:RMHMC}, and Shadow Hamiltonian Monte Carlo (SHMC) \citep{Izaguirre2004}, discussed in \S \ref{sec:Shadow_HMC}. We will conclude this review with an outline of the most recent research direction in \S \ref{sec:recent_advances}, including stochastic gradient methods and HMC in infinite-dimensional spaces.
 

\section{GEOMETRY AND PHYSICS} \label{sec:geometry_physics}

In HMC the sample space $\mathcal{X}$ is viewed as a (possibly high-dimensional) space called a manifold, over which a motion is imposed. The reader should keep in mind the idea of a fluid particle moving on the sample space (here, the manifold - see Fig. \ref{fig:simulation_HMC_RWM}-b.); the algorithm proposes new states by following the trajectory of this particle for a fixed amount of time. By coupling the choice of Hamiltonian dynamics to the target density, the new proposals will allow us to explore the density more efficiently by reducing the correlation between samples, and hence make MCMC more efficient. This paper seeks to explain why this is the case.

In this section we provide an accessible introduction to notions of geometry that are required to define Hamiltonian mechanics. Our hope is to provide the bare minimum of geometry in order to provide some insight into the behaviour of the Markov chains obtained. The avid reader is referred to \cite{Arnold1989,Frankel2012} for a more thorough introduction to geometry and physics, and to \cite{Amari1987,Murray1993} for the interplay of geometry and statistics. In particular, some of the concepts presented here also have a role in the study of statistical estimation, shape analysis, probability distributions on manifolds and point processes \citep{Kass1997,Dryden1998,Chiu2013,Dryden2015}.

\subsection{Manifolds and Differential Forms} \label{sec:manifolds_differential_forms}

Manifolds generalise the notions of smooth curves and surfaces to higher dimensions and are at the core of modern mathematics and physics. Simple examples include planes, spheres and cylinders, but more abstract examples also include parametric families of statistical models. Manifolds arise by noticing that smooth geometrical shapes and physical systems are coordinate-independent concepts; hence their definitions should not rely on any particular coordinate system. Coordinate patches (defined below) assign coordinates to subsets of the manifold and allow us to turn geometric questions into algebraic ones. In particular, the coordinate patches allow us to transfer the calculus of $\mathbb{R}^d$ to the manifold. Note that it is rarely possible to define a single coordinate patch over the entire manifold, except for the simplest manifolds.

\begin{figure}[t]
\includegraphics[width=0.7\textwidth,clip,trim = 0 0 0 0]{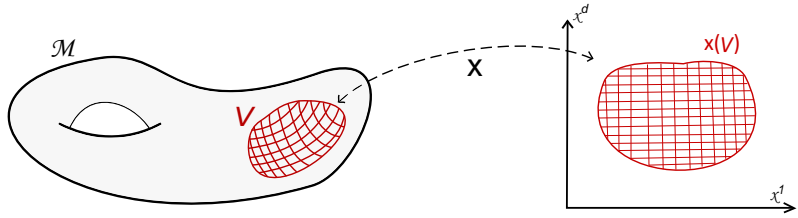}
\caption{The coordinate patch $\bm x$ attaches coordinates $(x^1,\ldots,x^d)$ to points in the neighbourhood $V\subseteq \mathcal M$.}
\label{fig:coordinate_patches}
\vspace{-4mm}
\end{figure}

A \(d\)-\textbf{dimensional manifold} is a set $\mathcal{M}$ such that every point $q\in \mathcal{M}$ has a neighbourhood $V\subseteq \mathcal{M}$ that can be described by $d$-coordinate functions $(x^1,\ldots,x^d)$. This means that there exists a bijection $\bm x_V:V\rightarrow \bm x_V(V) \subseteq\mathbb{R}^d$, called a \textbf{coordinate patch}, which assigns the coordinates $\bm x_V(q)=\big(x^1(q),\ldots,x^d(q)\big)$ to $q$. The functions $x^j :V\rightarrow \mathbb R$ are called \textbf{local coordinates}, we view these coordinates as being imprinted on the manifold itself (see Fig. \ref{fig:coordinate_patches}). 
Whenever two patches $\bm x_{V},\bm x_{W}$ overlap,  $V \bigcap W \neq \emptyset$, any point $q$ in the overlap is assigned two coordinates $\bm x_W(q),\bm x_V(q)$; in this case we require the patches to be \textbf{compatible}, i.e., the map $\bm x _V \circ \bm x_W^{-1}$, which is just the map that relates the coordinates should be smooth ($C^{\infty}$).

\begin{marginnote}
\textbf{Manifolds} \\
Technically, for $\mathcal M$ to be a manifold we further require that the topology generated by the differential structure consisting of all compatible patches be Hausdorff and have a countable base. See \citep{Arnold1989} for more details.
\end{marginnote}

\begin{figure}
\begin{subfigure}[b]{0.24\textwidth}
\includegraphics[width=\textwidth,clip,trim = 0 0 0 0]{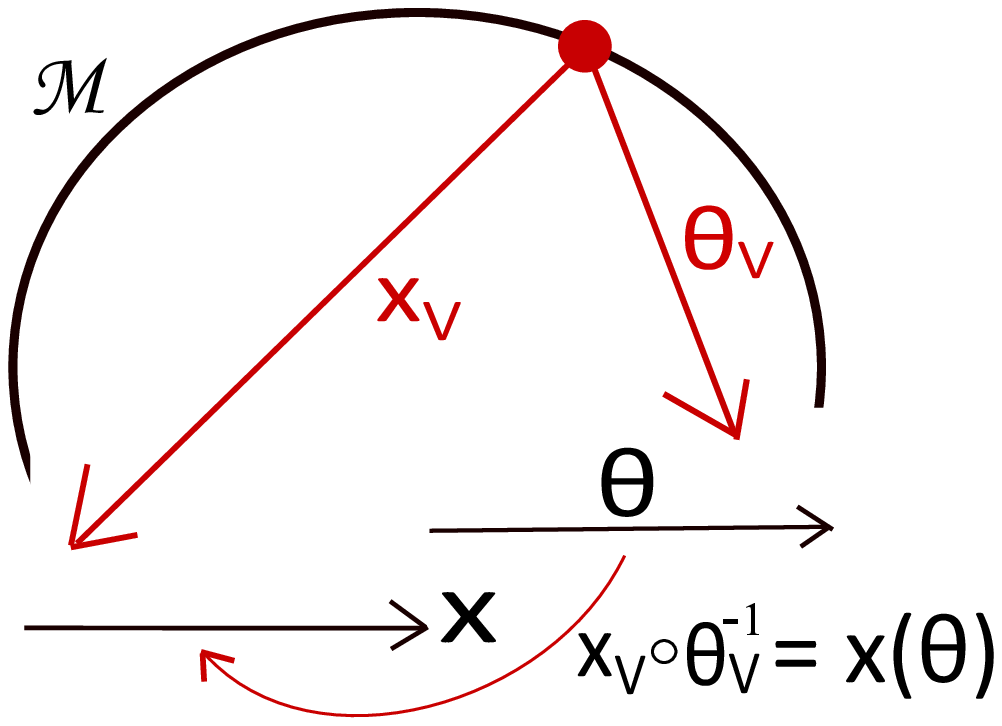} 
\caption{ }
\end{subfigure}
\begin{subfigure}[b]{0.24\textwidth}
\includegraphics[width=0.9\textwidth,clip,trim = 0 0 0 0]{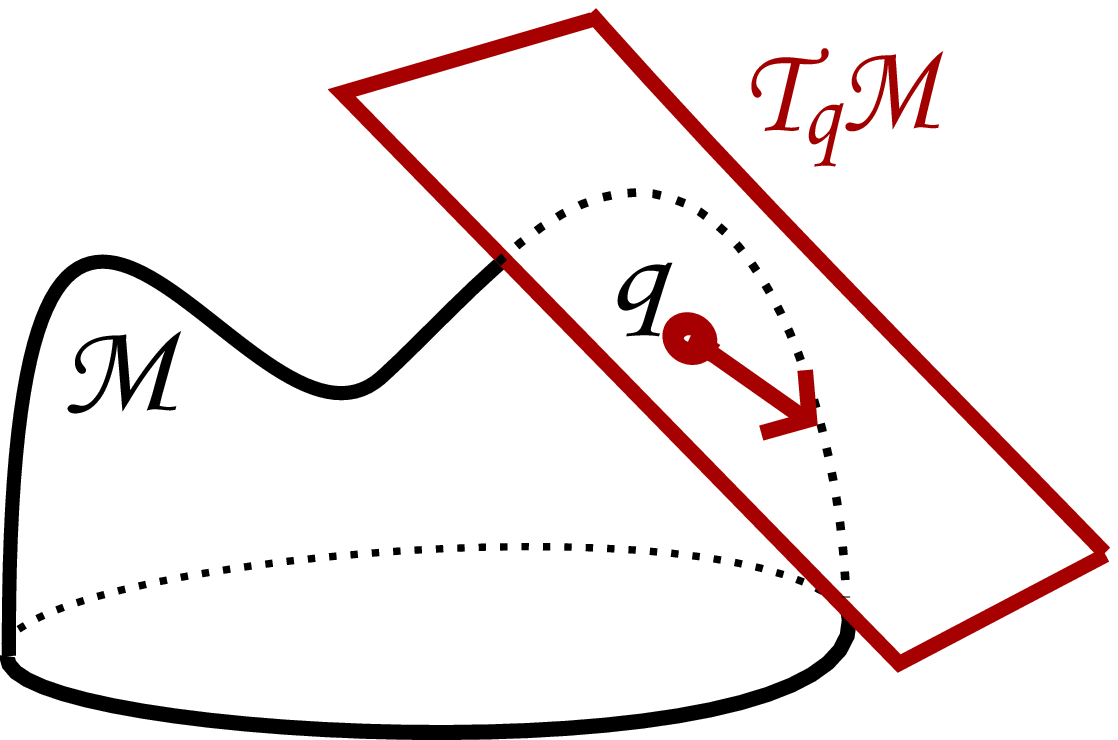}
\caption{ }
\end{subfigure}
\begin{subfigure}[b]{0.24\textwidth}
\includegraphics[width=0.7\textwidth,clip,trim = 0 0 0 0]{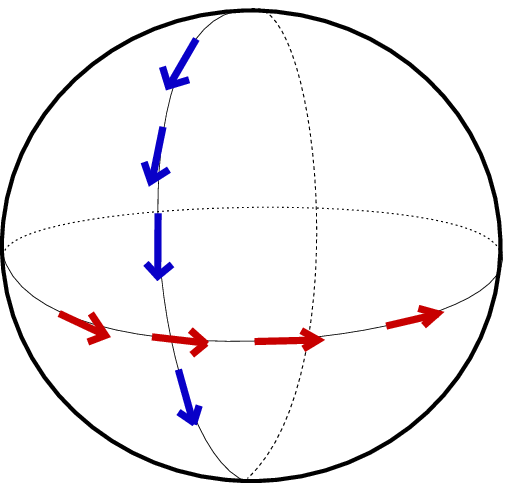}
\caption{ }
\end{subfigure}
\begin{subfigure}[b]{0.24\textwidth}
\includegraphics[width=\textwidth,clip,trim = 0 0 0 0]{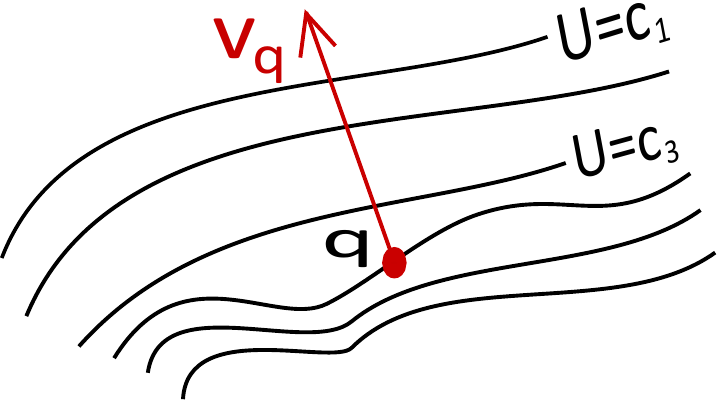}
\caption{}
\end{subfigure}
\vspace{3mm}
\caption{Manifolds and differential forms - \textbf{(a)} Patches $\bm x_V,\bm \theta_V$ on the upper hemisphere assign different real numbers to point on the circle $\mathcal M=S^1$. \textbf{(b)} Tangent vectors at $q$ belong  to the tangent space $T_q\mathcal M$. \textbf{(c)} On a sphere with coordinates $(\theta,\varphi)$, $\theta\in (0,2\pi)$, $\varphi \in (0,\pi)$, the vector field $\partial_{\theta}$ (in red) is tangent to the $\theta$-coordinate lines (lines of constant $\varphi$). $\partial_{\varphi}$ (in blue) is tangent to the $\varphi$-coordinate lines. \textbf{(d)} The 1-form $d_qU$ when applied to $v_q$ tells us how much potential $U$ is gained along the vector $v_q$.}
\label{fig:manifolds_diff_forms}
\end{figure}

For example, two possible patches for the (1-dimensional) semi circle $x^2+y^2=1$, $y>0$ in a neighbourhood of $(0,1)$ are $\bm x_V\big{(}(x,y)\big{)}=x$, and $\bm \theta_V \big{(}(x,y)\big{)}=\theta$ where $\theta$ satisfies $(\cos\theta,\sin \theta)=(x,y)$. Note the smoothness of $\bm x_V \circ \bm \theta^{-1}_V=x(\theta)=\cos \theta$, $ \bm \theta_V \circ \bm x^{-1}_V=\theta(x)=\cos^{-1}x$ implies the patches are compatible (see Fig. \ref{fig:manifolds_diff_forms}-a.). 
The sphere $S^2$ is a 2-(sub)manifold in $\mathbb{R}^{3}$. In a neighbourhood of the north pole, points are specified by their $x,y$ coordinates since we can write $z$ as the graph $z=z(x,y)=\sqrt{1-x^2-y^2}$. These points may be written as $\big{(}x,y,z(x,y)\big{)}$ and we can define a patch $\bm x\big{(}x,y,z(x,y)\big{)}=(x,y)$. Note we could have also used the spherical coordinates $(\theta, \varphi)$ on the upper half of $S^2$.

A more interesting example is that of the statistical manifold of Gaussian distributions $\mathcal M=\{\mathcal{N}(\cdot |\mu, \sigma^2): \mu \in \mathbb{R}, \sigma^2 \in \mathbb{R}_+\}$ which is a manifold endowed with global coordinates $\bm x_{\mathcal M}\big{(}\mathcal{N}(\cdot|\mu, \sigma^2)\big)=(\mu,\sigma^2)$.

A function $f:\mathcal M\rightarrow \mathbb{R}$ on the manifold is said to be \textbf{smooth} at a point $q\in \mathcal M$ if there exists a coordinate patch $\bm x_V$ around $q$ such that $f_V:=f\circ \bm x_V^{-1}:\bm x_V(V)\rightarrow \mathbb{R}$ is smooth.\begin{marginnote}[]
{\textbf{Smooth function on the circle:} \\ If $f:S^1\rightarrow \mathbb R$, locally around $(0,1)$ $f\circ \bm x^{-1}_V=f(x)$ while $f\circ \bm \theta^{-1}_V=f(\theta)$
}
\end{marginnote}
Note the map $f_V$ is just the coordinate expression of $f$. Since the coordinate patches are compatible, this definition of smoothness is independent of the choice of patches. The space of smooth functions on $\mathcal M$ is denoted $C^{\infty}(\mathcal M)$.

To define Hamiltonian dynamics, we now introduce the concept of velocity of the flow of a particle on $\mathcal{M}$ (i.e. our sample space) defined by tangent vectors to the manifold. Recall that in $\mathbb{R}^d$, any vector $v=(v^1,\ldots,v^d)$ defines a directional derivative that acts on functions $f\in C^{\infty}(\mathbb{R}^d)$, by
$v(f) \; := \;
\nabla_vf \;= \;
v \cdot \nabla f \; = \;
\sum_{j=1}^d v^j\partial_j f$,
where $\partial_i:=\frac{\partial}{\partial x^i}$. We can thus think of the vector $v$ as a first order differential operator $v=\sum_{j=1}^d v^j\partial_j:C^{\infty}(\mathbb{R}^d)\rightarrow \mathbb{R}$ (which is linear and satisfies Leibniz rule). We now generalise this to manifolds:
if $f,h\in C^{\infty}(\mathcal M)$, we define a \textbf{tangent vector} $v_q:C^{\infty}(\mathcal M)\rightarrow \mathbb{R}$ at $q\in \mathcal{M}$ to be a linear map satisfying Leibniz rule.
\begin{marginnote}
\textbf{Leibniz rule} \\
Given a vector $v_q$ at some point $q \in \mathcal{M}$, Leibniz rules is given by: $v_q(fh)  =  f(q)v_q(h)+h(q)v_q(f)$
\end{marginnote}

Defining a linear combination of tangent vectors by $(au_q+bv_q)f:=au_q(f)+bv_q(f)$, turns the set of tangent vectors at $q \in \mathcal{M}$ into a vector space denoted $T_q \mathcal M$, called the \textbf{tangent space} at $q$ (see Fig.\ref{fig:manifolds_diff_forms}-b.). Consider a local coordinate patch $(V,\phi_V)$ around $q$. The coordinate functions $x^j$ define tangent vectors $\partial_j|_q$ at $q$ by 
\begin{equation*}
\partial_j|_q(f) \; := \; \left. \frac{\partial f_V}{\partial x^j}\right |_{\bm x(q)}.
\end{equation*}
These tangent vectors form a basis of $T_q\mathcal M$;  any tangent vector at a point $q$ is of the form $\sum_{j=1}^d v^j \partial_j|_q$, where $v^j\in \mathbb{R}$.  A \textbf{vector field} $v$ is a smooth map that assigns at each point $q$ a tangent vector $v_q$. Locally any vector field can be written as $v=\sum_{j=1}^d v^j(\bm x) \partial_j|_{\bm x}$ where $\partial_j$ is the (local) vector field $\partial_j:q\mapsto \partial_j|_q$.  See Fig. \ref{fig:manifolds_diff_forms}-c. for an example on the sphere.

The objects $df$ or $dx$ are often introduced as being mysterious ``infinitesimal vectors/quantities" that give a real number when integrated. These objects are in fact special cases of \textbf{differential forms} that we shall now formally introduce: they play a central role in Hamiltonian mechanics.

A \textbf{1-form} at a point $q\in \mathcal M$ is a linear functional on the tangent space, i.e., a linear map $\alpha_q:T_q\mathcal M\rightarrow \mathbb{R}$. The simplest example is the differential of a function, $d_qf$, which maps a vector $v_q$ to the rate of change of $f$ in direction $v_q$: $d_qf(v_q):=v_q(f)$. In a coordinate patch, we can consider the differential of the coordinate function $x^i$. Taking $v_q=\partial_j|_q$, we see that $d_qx^i(\partial_j|_q)=\partial_j|_q(x^i)=\delta_j^i$ where  $\delta_j^i$ is $1$ if $i=j$ and $0$ otherwise. 

\begin{marginnote}[]
\textbf{Example of Differential:} \\
{Let $(\theta,z)$ be coordinates on a cylinder. Suppose $f_V(\theta,z):=z^2-\theta$, then $df=2zdz-d\theta$. At $q=(1,3)$, $d_qf=6d_qz-d_q\theta$  }
\end{marginnote}

This shows that $(d_qx^j)$ is the dual basis to $(\partial_j|_q)$ and thus a basis of $T_q^{\ast}\mathcal M$, the vector space of 1-forms at $q$. A \textbf{differential 1-form} $\alpha $ is a smooth map that assigns at each point $q$ a 1-form $\alpha_q$. Locally (i.e., in a given coordinate patch) any differential 1-form may be written as $\alpha  =\sum_{j=1}^d \alpha _j(\bm x) dx^j$ where $d{x^j}$ is the (local) differential 1-form $dx^j:q\mapsto d_qx^j$. For example the differential of the function $f$ is
$df \; = \; \frac{\partial f_V}{\partial x^1} dx^1+\cdots +\frac{\partial f_V}{\partial x^d} dx^d.$

A physical example of a $1$-form is the force $F$ acting on a particle, which is given by the differential of a potential energy function $F=-dU$. In HMC, the potential energy $U$ is related to the target unnormalised density by $U:=-\log(\tilde{\pi})$.
Given a vector, the force measures the rate at which potential energy is gained by moving in that direction (see Fig. \ref{fig:manifolds_diff_forms}-d.). Directions of increasing $U$ correspond to directions of decreasing probability. 

\begin{marginnote}[]
\textbf{Length of Curves:} \\
{The inner product defines a norm $|| v ||^2 =g(v,v)$, the length of a curve $\gamma$ is given by integrating its tangent vector $\int_{\gamma}|| \dot{\gamma} ||$}
\end{marginnote}

Finally, to define the notions of volume, curvature and of length of curves on a manifold, it suffices to define the length of tangent vectors. A \textbf{Riemannian metric} $g$ is a smooth assignment of an inner product $g_q: T_q \mathcal M \times T_q\mathcal M\rightarrow \mathbb{R}$ at each point $q \in \mathcal{M}$. The pair $(\mathcal M,g)$ is called a \textbf{Riemannian manifold}. Sub-manifolds of $\mathbb{R}^d$ have a natural Riemannian metric which arises by simply restricting the standard inner product of $\mathbb{R}^d$ to the sub-manifold. In local coordinates we can define at each point $q$ a symmetric matrix $\bm{G}$ such that $\bm{G}_{ij}:=g_q(\partial_i|_q,\partial_j|_q)$. We then recover the usual inner product space result $g_q(v,u)=\bm{v}^T\bm{G}\big(\bm x(q)\big)\bm{u}$, where $\bm u$ is the array $(u^1,\ldots,u^d)$ of coefficients of the vector $u$ in the local coordinate basis $u = u^1\partial_1 + \cdots + u^d\partial_d $.

This now concludes our brief introduction to differential geometry. The tools developed above allow us to formalise Hamiltonian dynamics on manifolds, which will be used to create efficient proposals for MCMC.

\begin{marginnote}[]
\textbf{Historical Note:} \\
{Riemannian geometry was introduced in statistics by Rao, who noted the Fisher-Rao metric defined a useful notion of distance between populations.}
\end{marginnote}


\subsection{Hamiltonian Mechanics}\label{sec:Hamiltonian_dynamics}

\begin{figure}
\begin{center}
\begin{subfigure}[b]{0.3\textwidth}
\includegraphics[width=0.9\textwidth,clip,trim = 0 0 0 0]{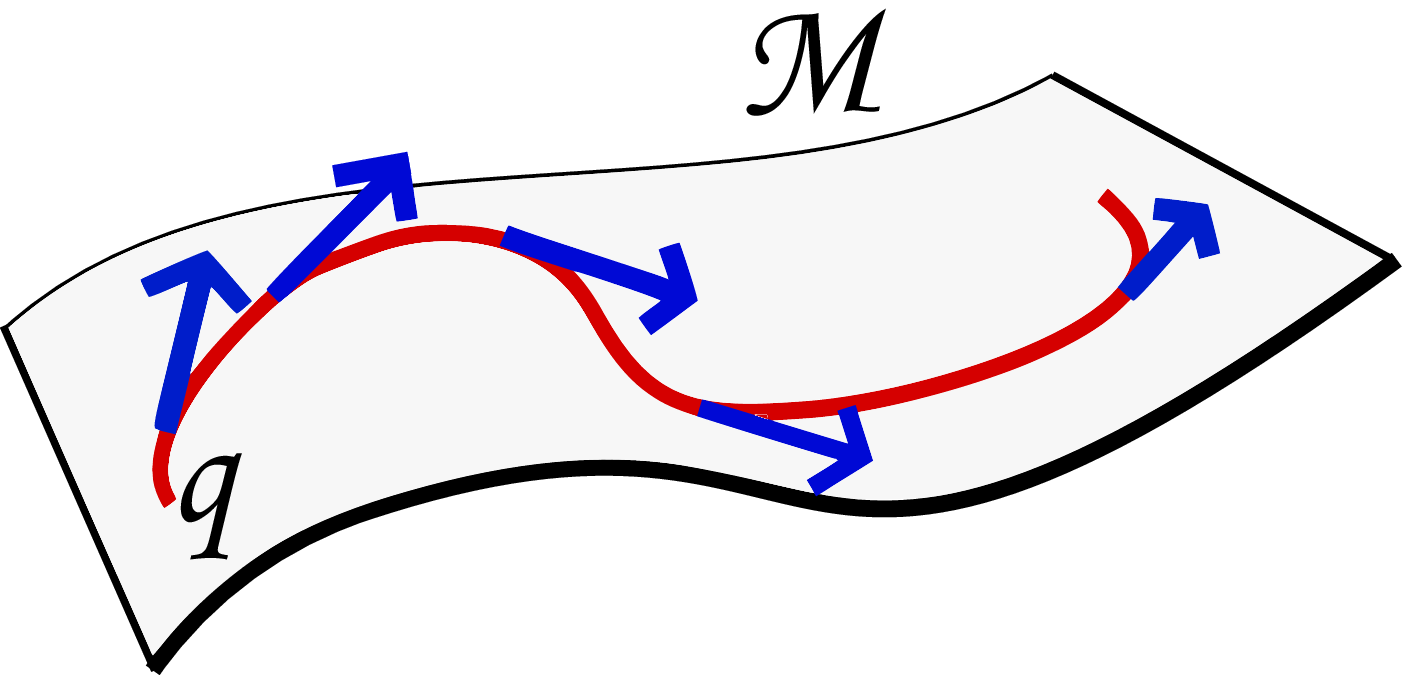}
\caption{}
\end{subfigure}
\begin{subfigure}[b]{0.3\textwidth}
\begin{center}
\includegraphics[width=0.6\textwidth,clip,trim = 0 0 0 0]{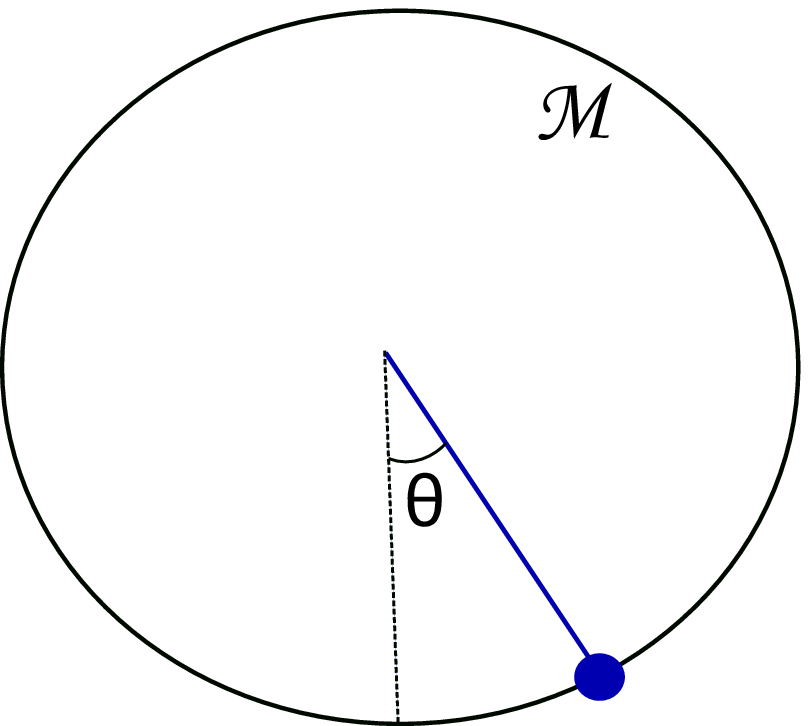}
\end{center}
\caption{}
\end{subfigure}
\begin{subfigure}[b]{0.3\textwidth}
\includegraphics[width=0.85\textwidth,clip,trim = 0 0 0 0]{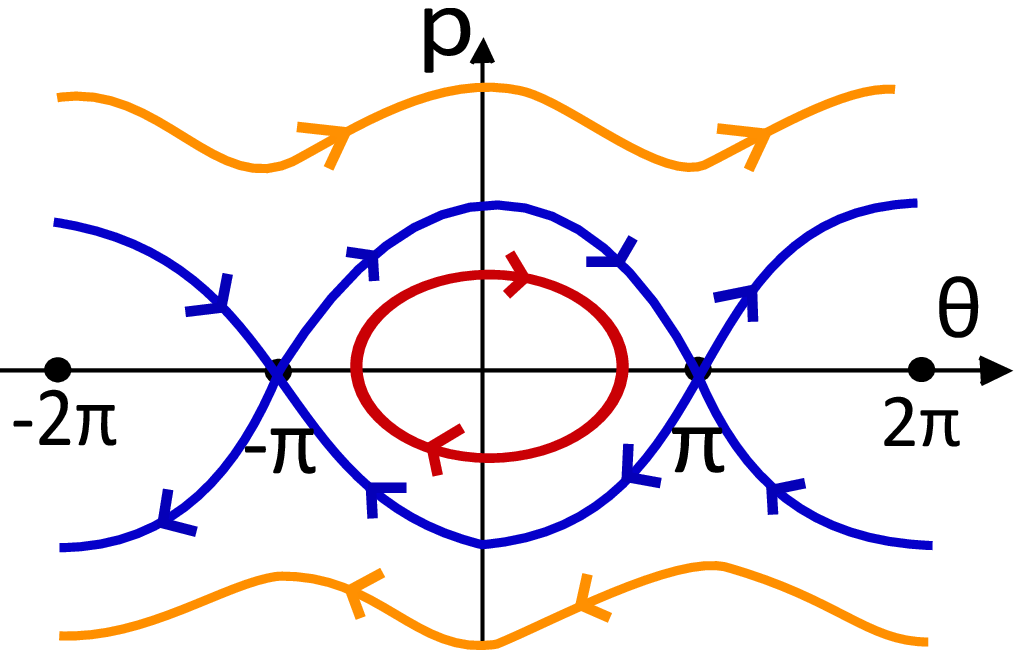}
\caption{}
\end{subfigure}
\end{center}
\vspace{2mm}
\caption{\textbf{(a)} The particle initially at $q$ follows the trajectory $\gamma=\Phi(q)$ (in red). Its velocity field $\dot{\gamma}$ is in blue. \textbf{(b)} A mass (in blue) attached to a pendulum has $\mathcal M=S^1$. As the mass moves from $\theta=0$ to $\theta= \pi$, its kinetic energy is transformed into potential energy. \textbf{(c)} Possible trajectories in phase space $\mathcal{N}$ for the mass pendulum. The red/blue/orange trajectories represent respectively the cases when there is not enough/exactly enough/more than enough energy to do a full turn.
}
\label{fig:Hamiltonian_Mechanics}
\end{figure}

Consider a particle moving on $\mathcal M$ from initial position $q\in \mathcal M$.  We call $\mathcal M$ the \textbf{configuration manifold} (or configuration space).  The particle could, for example, be a mass attached at the end of a plane pendulum (so $\mathcal M=S^1$) or a fluid particle flowing along a river. 
The deterministic motion followed by the particle is governed by the laws of physics. Let $\Phi_t(q)$ be its position at time $t$, so $\Phi_0(q)=q$, and the trajectory followed by the particle is given by the curve $\gamma:t \mapsto \Phi_t(q)$. The curve $\gamma$  generates a vector field $\dot{\gamma}$ over the range of $\gamma$ representing the velocity of the particle; the tangent vector at the point $\gamma(a)=r$ is defined, for any function $f$, by
\begin{equation*}
\dot{\gamma}_{r}(f) \; := \; \frac{df(\gamma(t))}{dt}\Big|_{t=a}= (f\circ\gamma)'(a).
\end{equation*}
Since the laws of physics are the same at all times, we have that $\Phi_{t}\circ \Phi_{s}(q)=\Phi_{t+s}(q)$. We call $\Phi$ the \textbf{flow} and $\dot{\gamma}$ the \textbf{velocity field} (see Fig.\ref{fig:Hamiltonian_Mechanics}-a.). 

The particle has a kinetic energy $K$ that measures the energy carried by its speed and mass. If no forces are acting, the particle's kinetic energy (and speed) will be constant; otherwise the force $F$ will increase/decrease the particle's kinetic energy. Since energy is conserved, the particle must be losing/gaining some other type of energy introduced by the force field, which we call potential energy $U$  (see Fig.\ref{fig:Hamiltonian_Mechanics}-b. for an example on the pendulum). It can be shown that $F=-dU$, which shows that the force is caused by variations in potential energy. \begin{marginnote}[]{
\textbf{Lagrangian:}\\
In general $p := \partial \mathcal L/\partial v$  where $\mathcal L(x, v)$ is the Lagrangian (see supplementary material). If $\mathcal L = K - U$ with $K = g(v,v)/2 = g(\dot\gamma,\dot\gamma)/2$ and if $g$ is constant then $p = g(\dot\gamma,\cdot)$; but this is not true in general.

}
\end{marginnote}

A Riemannian metric provides an identification between vector fields and differential 1-forms, by associating the vector field $v$ to the 1-form $\alpha(\cdot):=g(v,\cdot)$, and the inner product on vectors $g(u,v)=\bm u^T \bm G \bm v$ induces an inner product on the associated 1-forms (iff $\det \bm G\neq0$) by $g^{-1}(p,\alpha):=\bm p^T \bm G^{-1} \bm \alpha$.\begin{marginnote}[]{\textbf{Example of momentum field:} \\ Suppose a particle in a plane has momentum field $p=ye^xdx-xdy$. Then $\bm z=(x,y,ye^x,-x)$. At $q=(1,3)$ its momentum is $p_q=3ed_qx-d_qy$ and its phase is $\big (1,3,3e,-1\big )$.}
\end{marginnote}In particular each velocity field $\dot{\gamma}$ induced by a curve $\gamma$ has an associated \textbf{momentum field} defined by $p:=g(\dot{\gamma},\cdot)$ which represents the ``quantity of motion" in direction $\dot{\gamma}$. Writing $p=\sum_{j=1}^d p_j(\bm x) dx^j$ makes it clear that to define $p$ (i.e. to specify each 1-form $p_q$, called \textbf{momentum}, defined by $p$ at $q\in \mathcal{M}$) we need to specify the $2d$-tuple:$\big(\bm x(q), \bm p(q)\big) \; := \;\big(x^1(q),\ldots,x^d(q),p_1(q),\ldots,p_d(q)\big )$, i.e. the position $\bm x(q)$ of $p_q$ and the momentum components at $\bm p(q)$ .

\begin{marginnote}[]{
\textbf{Bundles:} locally a cartesian product of manifolds, but globally may be ``twisted" like a M\"obius strip.}
\end{marginnote}

This shows that the set of momenta (or equivalently the set of 1-forms) $T^{\ast}\mathcal M=\bigcup_q T_q^{\ast}\mathcal M$ is a $2d$-dimensional manifold, called the \textbf{cotangent bundle}, on which $\bm z :=(\bm x, \bm p)=(x^1,\ldots,x^d,p_1,\ldots,p_d)$ are coordinates.
 
 At any given time $t$, the 2$d$-tuple $\bm z\big(\gamma(t)\big)$, consisting of the position $\bm x\big(\gamma(t)\big )$ of the particle and its momentum $\bm p\big(\gamma(t)\big)$,
is called the \textbf{phase} and fully specifies the physical system, i.e., it encodes all the information about the system and determines its future dynamics. The space of all possible phases is called \textbf{phase space} or the cotangent bundle $T^{\ast}\mathcal M$ (see Fig.\ref{fig:Hamiltonian_Mechanics}-c. for the phase space of the pendulum example).

\begin{figure}[t]
\begin{subfigure}[b]{0.4\textwidth}
\includegraphics[width=0.9\textwidth,clip,trim = 0 0 0 0]{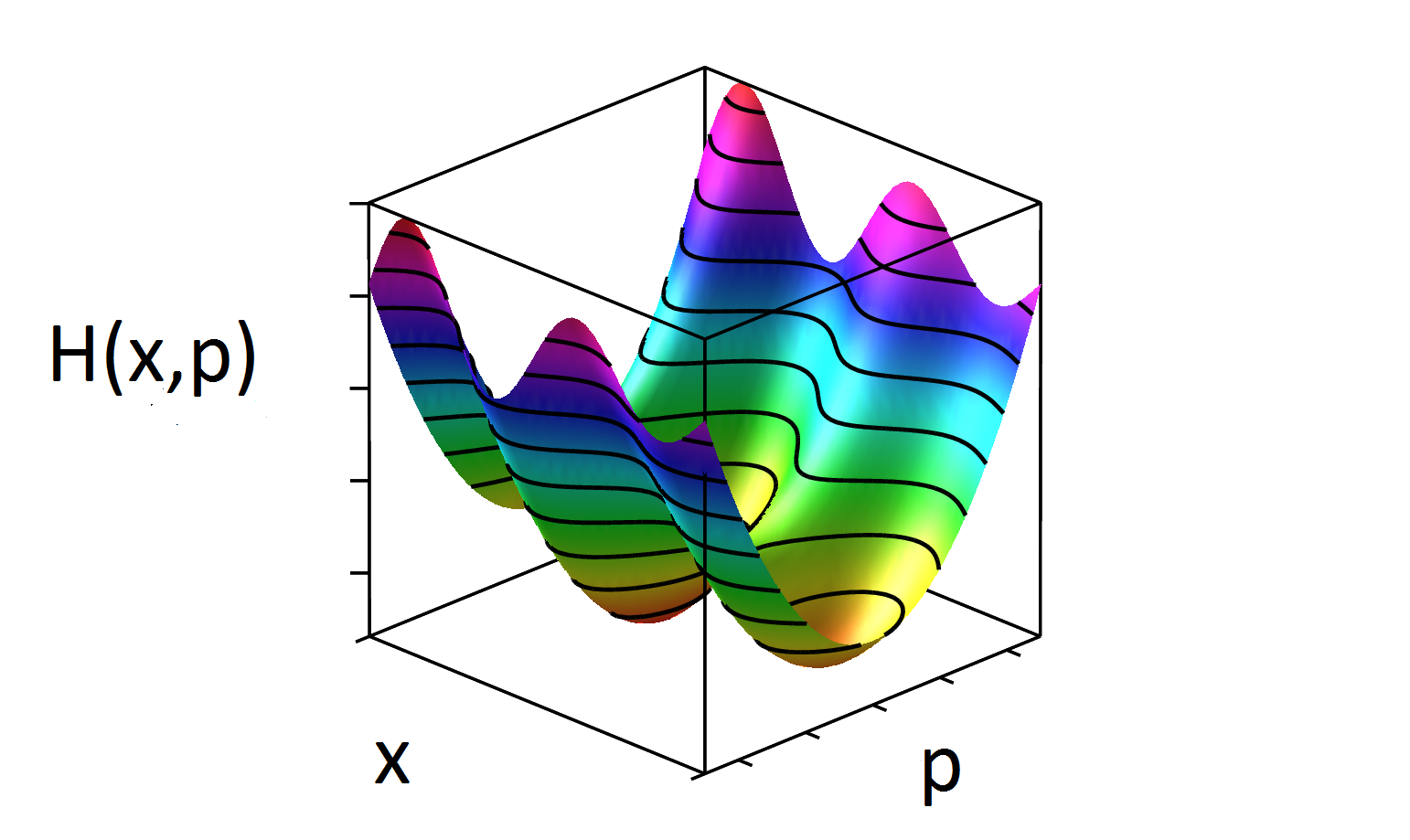}
\caption{}
\end{subfigure}
\begin{subfigure}[b]{0.29\textwidth}
\begin{center}
\includegraphics[width=0.75\textwidth,clip,trim = 0 0 0 0]{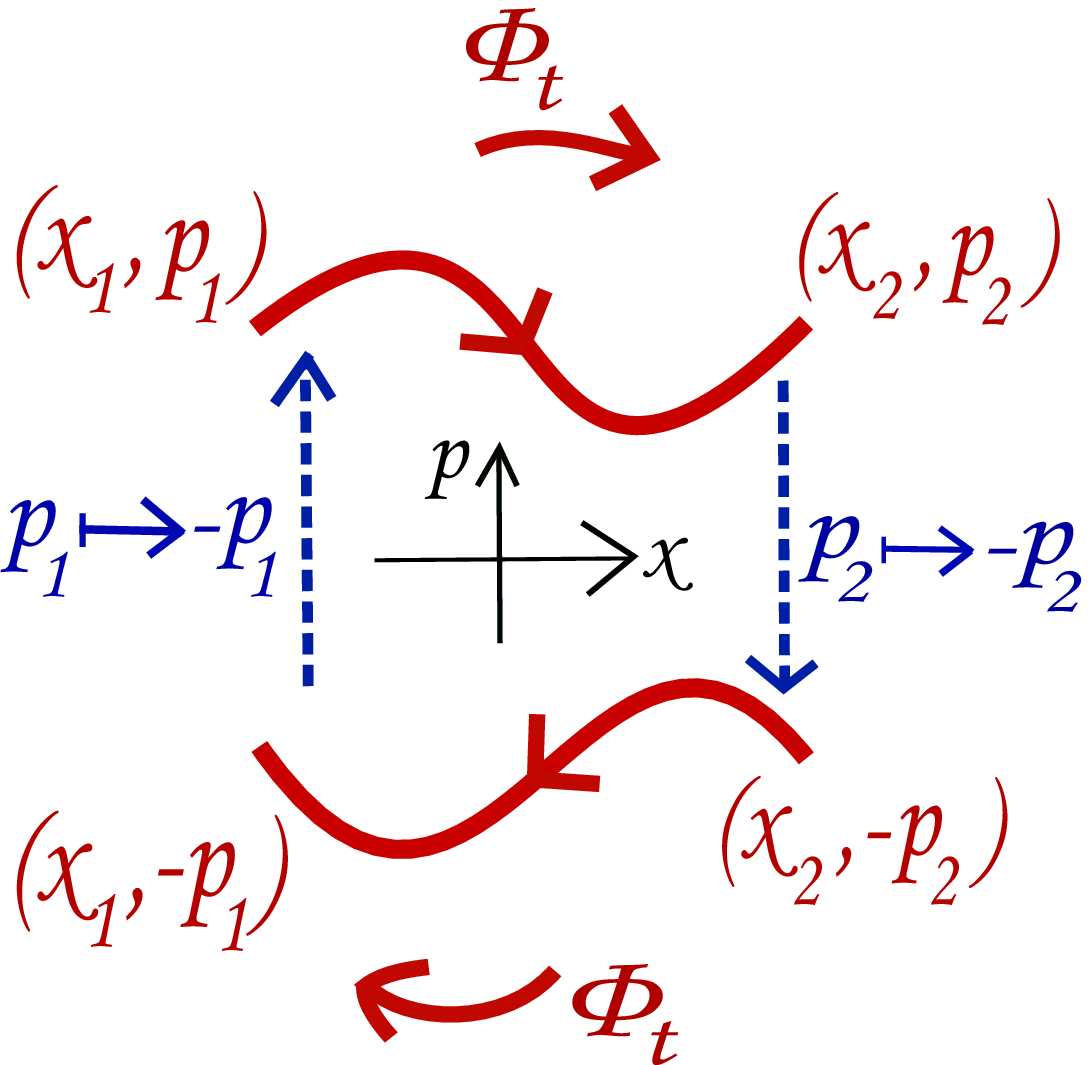}
\end{center}
\caption{}
\end{subfigure}
\begin{subfigure}[b]{0.29\textwidth}
\includegraphics[width=0.85\textwidth,clip,trim = 0 0 0 0]{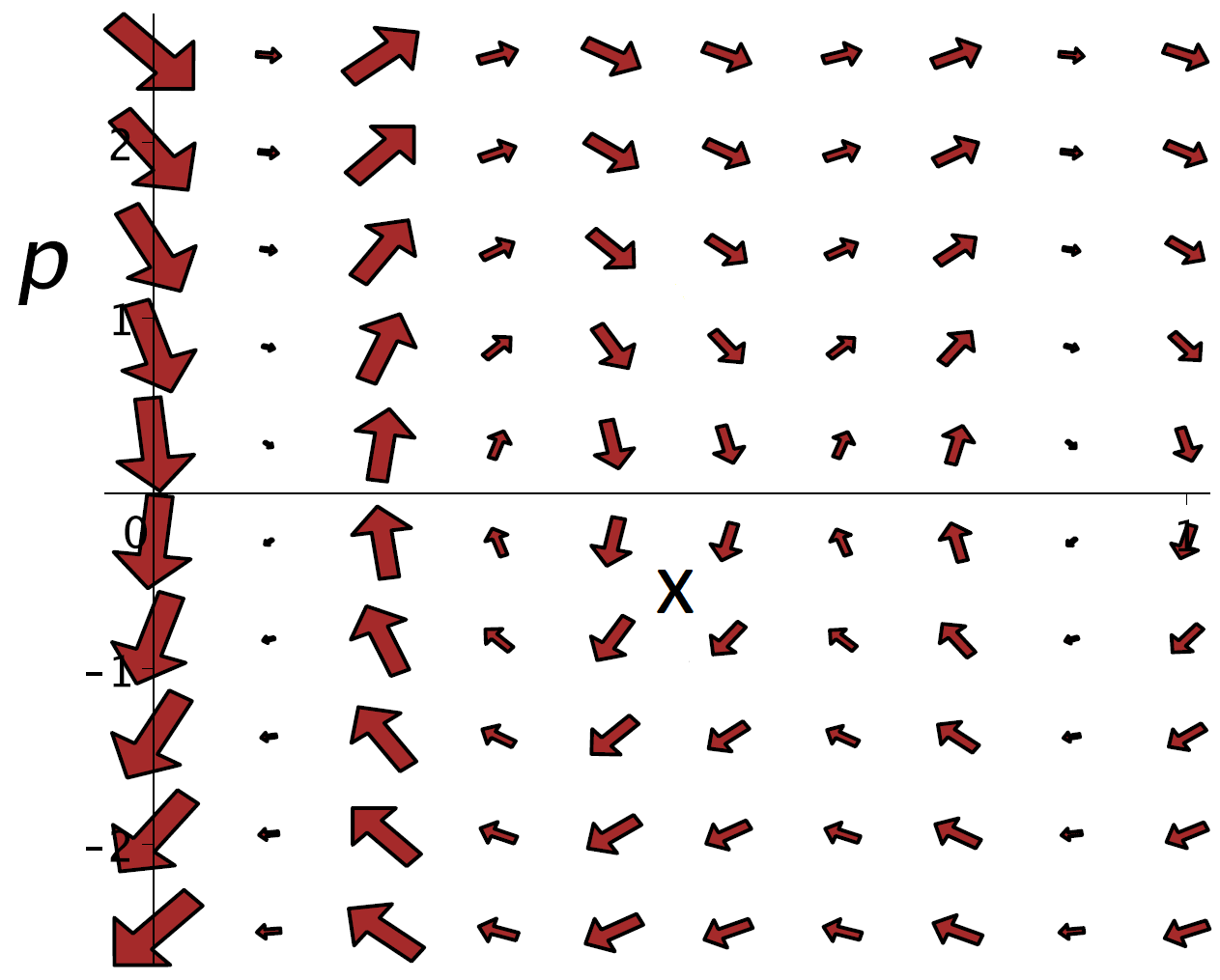}
\caption{}
\end{subfigure}
\vspace{2mm}

\caption{\textbf{(a)} Surface plot of a Hamiltonian with contour lines. \textbf{(b)} Time-reversibility of Hamiltonian mechanics. \textbf{(b)} Hamiltonian vector field for the system shown in (a).}
\label{fig:time_reversibility}
\end{figure}

We have seen earlier that forces acting on the system may be accounted by defining how the energy transfers between potential and kinetic. Hence, if we define the Hamiltonian function $H$ to be the total energy $H=K+U$, we expect its differential $dH$ to fully determine the dynamics of the system (from now on $K$ is a function of the momentum rather than the velocity), see Fig. \ref{fig:time_reversibility}-a.
We now construct Hamiltonian mechanics, in which the trajectory of the particle on $\mathcal M$ is described by a trajectory in phase space $\mathcal N:=T^{\ast}\mathcal M$ defining how the phase of the system evolves. From here on $\Phi$ is a flow on $\mathcal N$ and $\gamma$ a curve $t \mapsto \Phi_t(\bm z_0)$ on $\mathcal N$ for some initial phase $\bm z_0$ \big(locally $\gamma(t)$ is now described by coordinates $\bm z(t):=\big (\bm x(\gamma(t)), \bm p(\gamma(t)\big)\big)$ and $\bm x\big(\gamma(t)\big)$ are the coordinates of the physical trajectory in $\mathcal M$\big). To do so we will need a map that turns $dH$ into a trajectory $\gamma$ that is consistent with the laws of physics. This map is called a \textbf{symplectic 2-form} and we now proceed to describe it.

\begin{marginnote}
\textbf{Bilinear map:}
A map which is linear in each of its arguments.
\end{marginnote}

We need at each phase $\bm z\in \mathcal N$ an invertible linear map (since Newton's equations are linear) $S_{\bm z}^{-1}:T_{\bm z}^{\ast}\mathcal N \rightarrow T_{\bm z}\mathcal N $ to turn the differential form $dH$ into the vector field $\dot \gamma$ generated by the trajectory in phase space $\gamma$ (this vector field yields the velocity field when projected to the configuration space $\mathcal M$). Its inverse $S_{\bm z}:T_{\bm z}\mathcal N\rightarrow T_{\bm z}^{\ast}\mathcal N $ maps linearly vectors into 1-forms and fully determines Hamiltonian dynamics. 
Any such linear map $S_{\bm z}$ may be identified with a bilinear map $\omega_{\bm z}: T_{\bm z}\mathcal N \times T_{\bm z}\mathcal N  \rightarrow \mathbb{R}$ where $\omega_{\bm z}(u,v) = \big(S_{\bm z}(u)\big)(v)$. Letting $\omega$ be the smooth map $\bm z \mapsto \omega_{\bm z}$, note that since $S(\dot{\gamma})=dH$, then $\omega(\dot{\gamma},\cdot)=dH(\cdot)$, i.e., $\omega(\dot{\gamma},\cdot)$ maps a vector field to the rate of change of $H$ along it.
A \textbf{differential 2-form} $\beta$ is a smooth map that assigns to each $\bm z\in \mathcal N$ a bilinear, antisymmetric map $\beta_{\bm z}:T_{\bm z}\mathcal N \times T_{\bm z}\mathcal N \rightarrow \mathbb{R}$. We will now show that $\omega$ is a \textbf{symplectic 2-form} (also called symplectic structure), i.e., it satisfies: 
\begin{enumerate}
\item \textbf{Non-degenerate differential 2-form}: By the law of \textbf{conservation of energy}, the total energy of the system must be constant, $\frac{d}{dt}\big(H\circ \gamma(t)\big) \; = \; 0$ or equivalently $dH(\dot{\gamma})=0$. Thus, for all flows, we have $\omega(\dot{\gamma},\dot{\gamma})=0$, which implies\begin{marginnote}[]
\textbf{Symplectic 2-form} \\
{The symplectic 2-form $\omega$ turns $dH$ into a trajectory $\gamma$ through $\omega(\dot{\gamma},\cdot)=dH$. The properties of $\omega$ ensure $\gamma$ is compatible with physics}
\end{marginnote}$\omega$ is antisymmetric and thus a differential 2-form. Moreover $\omega$ is ``non-degenerate", which means the velocity field $\dot {\gamma}$ exists globally.
\vspace{1mm}
\item \textbf{Closed}: The laws of physics must be conserved in time, which mathematically means that $\omega$ is conserved along the flow and is ensured by demanding that its differential vanishes, $d\omega=0$ (the differential of a 2-form is formally defined in the supplementary material). This gives rise to \textbf{conservation of volume}: if particles are initially occupying a region $U$ in phase space with volume vol($U$), this volume will be preserved as they follow the flow, i.e., vol$\big{(}\Phi_t(U)\big{)}=$ vol($U$).
\end{enumerate}
\begin{marginnote}[]
\textbf{Example of wedge product:}\\
In $\mathbb{R}^3$, $dy \wedge dz$ applied to $(3,2,-5)$ and $(1,7,4)$ gives the signed area of parallelogram spanned by $(2,-5)$ and $(7,4)$.
\end{marginnote}
When $\mathcal M=\mathbb{R}^d$, the phase space $\mathcal N= \mathbb{R}^{2d}$ has a natural symplectic structure, which in (global) coordinates $(x^1,\ldots,x^d,p_1,\ldots,p_d)$ is given by
\begin{equation*}
\omega \; := \; dx^1\wedge dp_1+\cdots+dx^d\wedge dp_d. 
\end{equation*}
Here $d_{\bm z}x^i\wedge d_{\bm z}p_j:T_{\bm z}\mathcal N\times T_{\bm z}\mathcal N \rightarrow \mathbb{R}$ is the 2-form constructed using the wedge product that, given a pair of vectors, gives the signed area of the parallelogram spanned by their projection to the $x^i$--$p_j$ plane (see supplementary material).

The condition $\omega(\dot{\gamma},\cdot)=dH(\cdot)$ implies that the coordinate expression of $\gamma$, $\big(x^1(t),\ldots,x^d(t),p_1(t),\ldots,p_d(t)\big)$, satisfies \textbf{Hamilton's equations} 
\begin{equation}\label{eq:Hamiltonian_system}
\frac{dx^i}{dt} \; = \; \frac{\partial H}{\partial p_i}, \quad \quad 
\frac{dp_i}{dt} \; = \; -\frac{\partial H}{\partial x^i},
\end{equation}
i.e., the velocity field is orthogonal to the gradient of $H$. 
\begin{marginnote}
\textbf{Canonical Symplectic matrix:} \\
The canonical symplectic matrix is given by:
$$\bm J =
 \begin{pmatrix}
  \bm 0 &  I_{d\times d} \\
   -I_{d\times d} & \bm 0
 \end{pmatrix}$$ 
and we can rewrite Hamilton's equation as:$$\frac{d \bm z}{dt}= \bm J \frac{\partial H}{\partial \bm z}.$$.
\vspace{-6mm}
\end{marginnote}Finally, notice Hamiltonian mechanics is time reversible, i.e. Hamilton's equations are preserved under the transformation $t \rightarrow -t$, $\bm x \rightarrow \bm x$, $\bm p \rightarrow -\bm p$. This means the  following: consider a system, say a pendulum, with initial state $(\bm x_1,\bm p_1)$. After a time $t$ it will have a state $(\bm x_2,\bm p_2)=\Phi_t(\bm x_1,\bm p_1)$. If we reverse its momentum, $(\bm x_2,\bm p_2) \mapsto (\bm x_2,-\bm p_2)$, then after another time $t$ it will be at its initial position with opposite momentum, i.e., $\Phi_t\big{(} \bm x_2,-\bm p_2 \big{)}=(\bm x_1,-\bm p_1)$ (see Fig. \ref{fig:time_reversibility}-b.). \textbf{Time-reversibility} is necessary for detailed balance to hold in MCMC.

We have now defined the basic notions necessary to define Hamiltonian dynamics. More precisely, we have explained how the motion of a fluid particle on the manifold $\mathcal M$ is described by a curve in phase space $\mathcal{N} = T^{\ast}\mathcal M$. For this curve to represent a physical path, we have shown it must be related to the differential of the Hamiltonian $dH$ through a symplectic form $\omega$.


\section{HAMILTONIAN MONTE CARLO} \label{sec:Hamiltonian_Monte_Carlo}

In this section, we start by describing popular methods to approximate Hamiltonian dynamics on symplectic manifolds, then demonstrate how this can be used as a proposal within a Metropolis-Hastings algorithm.

\subsection{Hamiltonian Dynamics}

In practice Hamilton's equations cannot be solved exactly and we need to employ numerical methods that approximate the flow in Eq. \ref{eq:Hamiltonian_system} \citep{Leimkuhler2004,Hairer2006}. Let $\bm{z}(t_0):=(\bm{x}(t_0),\bm{p}(t_0))$ be the initial phase of a Hamiltonian system $H$. If we fix a time-step $\tau$ we can obtain a sequence of points along the trajectory that describe how the phase evolves
\begin{equation*}
\bm{z}(t_0) 
\; \rightarrow \;
\bm{z}(t_1) \; := \; \Phi_{\tau}(\bm{z}(t_0)) 
\; \rightarrow \;
\bm{z}(t_2) \; := \; \Phi_{\tau}(\bm{z}(t_1)) \; = \; \Phi_{\tau}^2(\bm{z}(t_0))
\; \rightarrow \;
\cdots 
\end{equation*}
A \textbf{numerical one-step method} is a map $\Psi_{\tau}$ that approximates this trajectory  
\begin{equation*}
\bm{z}(t_0) 
\; \rightarrow \;
\bm{z}^1 
\; := \;
\Psi_{\tau}(\bm{z}(t_0)) 
\; \rightarrow \;
\bm{z}^2 
\; := \; 
\Psi_{\tau}(\bm{z}^1) 
\; = \; 
\Psi_{\tau}^2(\bm{z}(t_0))
\; \rightarrow \; \ldots
\end{equation*}
where $\bm{z}^k=(\bm{x}^k,\bm{p}^k)$ approximates $\bm{z}(t_k)$. The numerical method will introduce an error at each step, defined as the difference between the application of $\Phi_{\tau}$ and $\Psi_{\tau}$ to a phase $\bm{z}$. Such errors will accumulate over time and the approximated trajectory will gradually deviate from the exact one. To partially remedy this we make use of geometric integrators which are numerical methods that exactly preserve some fundamental properties of the dynamics they simulate, and hence ensure the approximated trajectory retains some key features.

\begin{marginnote}
\textbf{Hamiltonian Mechanics:} \\
William Rowan Hamilton developed Hamiltonian mechanics as a generalisation of classical dynamics by applying ideas from optics and by re-formulating Lagrangian mechanics.
A more general introduction of Hamiltonian and Lagrangian dynamics is presented in the supplementary material. This may be of interest to readers interested in gaining a deeper understanding of some of the more advanced HMC methods.
\end{marginnote} 

 In particular \textbf{symplectic integrators} are geometric integrators that preserve the symplectic structure $\omega$ and thus the volume in phase space.

Any smooth map $S:\mathbb{R}^{2d}\rightarrow \mathbb{R}^{2d}$ has at each phase $\bm{z}=(\bm x,\bm p) \in T^*\mathcal{M}$ a Jacobian matrix $\bm{S}_{\bm{z}}$ which is a linear map $T_{\bm{z}}\mathbb{R}^{2d}\rightarrow T_{\bm{z}}\mathbb{R}^{2d}$. We say $S$ is a \textbf{symplectic map} if $
\bm{S}_{\bm{z}}^T\bm{J}^{-1}\bm{S}_{\bm{z}} \; = \; \bm{J}^{-1}$, where $J$ is the canonical symplectic matrix. The method $\Psi_{\tau}$ is called a \textbf{symplectic integrator} if it is a symplectic map. Writing $\bm{z}^{k+1}=\Psi_{\tau}(\bm{z}^k)$, this is equivalent to requiring that it preserves the symplectic structure for each step $k$:
\begin{equation*}
d\bm{x}^{k+1}\wedge d\bm{p}^{k+1}
\; = \; 
d\bm{x}^{k}\wedge d\bm{p}^{k}.
\end{equation*}

\begin{figure}
\begin{subfigure}[b]{0.41\textwidth}
\begin{center}
\includegraphics[width=0.9\textwidth,clip,trim = 0 0 0 0]{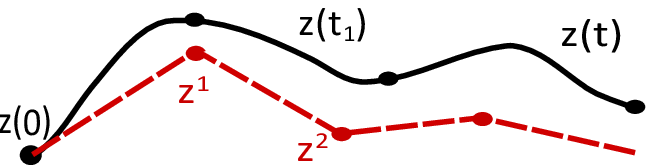}
\end{center}
\caption{}
\end{subfigure}
\begin{subfigure}[b]{0.27\textwidth}
\begin{center}
\includegraphics[width=\textwidth,clip,trim = 0 0 0 0]{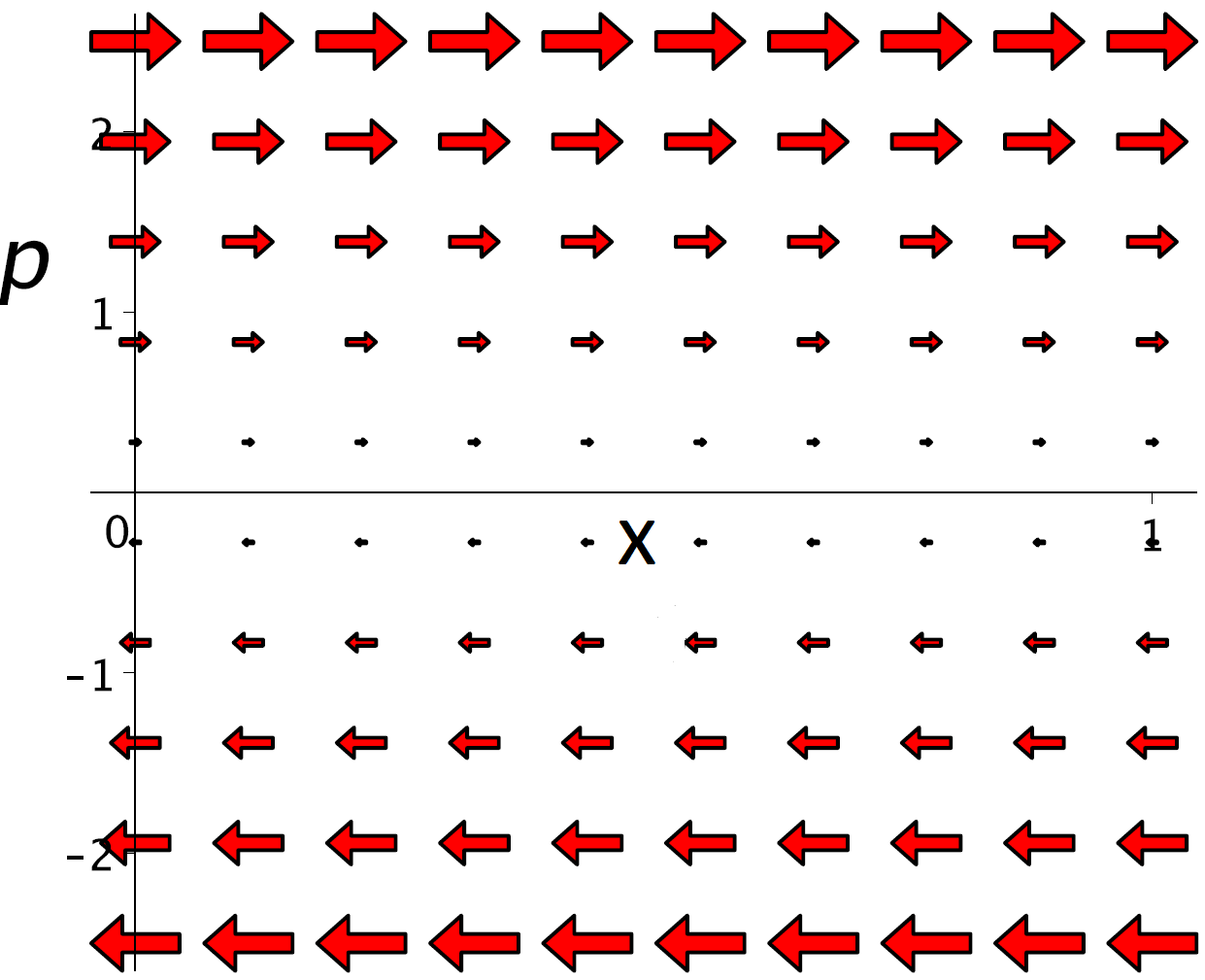}
\end{center}
\caption{}
\end{subfigure}
\begin{subfigure}[b]{0.27\textwidth}
\begin{center}
\includegraphics[width=\textwidth,clip,trim = 0 0 0 0]{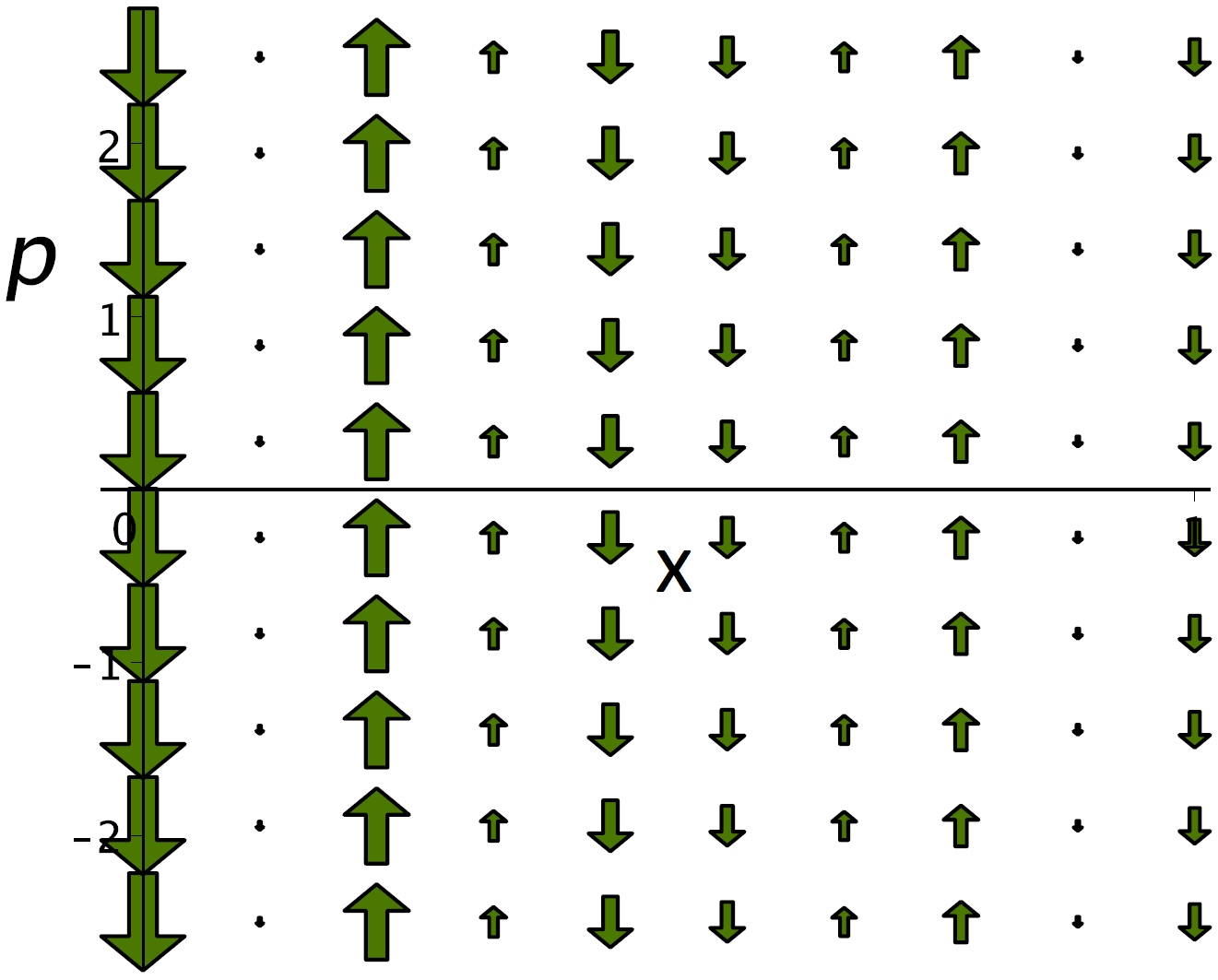}
\end{center}
\caption{}
\end{subfigure}
\vspace{3mm}
\caption{\textbf{(a)} Exact evolution of the phase $\bm{z}(t)$ (black curve) and numerical one-step method (red dashed curve). \textbf{(b--c)} Hamiltonian vector fields for $K$ and $U$ of the system in Fig. \ref{fig:time_reversibility}-a. respectively, both of which can be integrated exactly.}
\label{fig:simulating_Hamiltonian}
\end{figure}

A useful technique to easily build symplectic integrators uses Hamiltonian splitting.\begin{marginnote}[]
\textbf{Symplectic map:}\\
The definition given here is a local one. A symplectic map $S:\mathcal M \rightarrow \mathcal M$ is  one for which the induced map on 2-forms preserves $\omega$, $S_*:\omega\mapsto\omega$ (see supplementary material for definition of induced map).
\end{marginnote} Suppose our Hamiltonian is of the form $H=H_1+\cdots+H_{\ell}$ where Hamilton's equations may be solved explicitly for each Hamiltonian $H_i$. If we denote by $\Phi_{\tau}^{H_k}$ the exact flow of $H_k$ we can define a numerical method for $H$ by 
\begin{equation*}
\Psi_{\tau} \; := \; \Phi_{\tau}^{H_1}\circ \cdots \circ \Phi_{\tau}^{H_{\ell}}.
\end{equation*}
Note that the composition of these exact flows may not give the exact flow of $H$. However, since each flow $\Phi_{\tau}^{H_k}$ is symplectic, and the composition of symplectic maps is symplectic, $\Psi_{\tau}$ will be a symplectic integrator. The most popular symplectic integrator is the St\"ormer--Verlet or \textbf{leapfrog integrator} (see Fig. \ref{fig:simulating_Hamiltonian}-a.), which is derived through the splitting ((see Fig.\ref{fig:simulating_Hamiltonian}-b.) $H_1=\frac{1}{2}U(\bm{x})$, $H_2=K(\bm{p})$ and $H_3=\frac{1}{2}U(\bm{x})$ which gives 
\begin{eqnarray*}
\bm{p}^{k+\frac{1}{2}}
& = & \bm{p}^k-\frac\tau2 \frac{\partial U}{\partial \bm{x}}\big(\bm{x}^k \big),\\
\bm{x}^{k+1}
& = &
\bm{x}^k+\tau \frac{\partial K}{\partial p}\big(\bm{p}^{k+\frac{1}{2}}\big ), \\
\bm{p}^{k+1} 
& = & 
\bm{p}^{k+\frac{1}{2}}-\frac\tau2 \frac{\partial U}{\partial \bm{x}}\big(\bm{x}^{k+1}\big).
\end{eqnarray*}
It is easy to verify that the leapfrog integrator is reversible, i.e., we can invert the leapfrog trajectory by simply negating the momentum, applying the leapfrog algorithm and negating the momentum  again. It is also symmetric, $\Psi_{-\tau}^{-1}=\Psi_{\tau}$. Reversibility and conservation of volume of the integrator are required to prove detailed balance when we apply it in HMC. Note however that the energy is only approximately conserved along a leapfrog trajectory. 

The leapfrog integrator is an integrator of order 2 which means that its global error is of order $\tau^3$, where $\tau$ is the step-size. In situations in which very high accuracy is needed it may be necessary to turn to higher order integrators to obtain better approximations of the exact trajectory over a short time interval \citep{Campostrini1990,Yoshida1990,Leimkuhler2004}. The improved accuracy must however be balanced with the increased computational cost. Note that other integrators have also been proposed, see for example \cite{Blanes2014}.

\subsection{The Hamiltonian Monte Carlo Algorithm}

Suppose we want to sample from a probability density $\pi_{\bm x}:\mathcal{X}\rightarrow \mathbb R$ which we only know up to multiplicative constant: $\pi_{\bm x}=\tilde{\pi}_{\bm x}/Z$. The differential of $U(\bm{x}):=-\log\pi_{\bm x}(\bm{x})$, if it is known, is very useful as it informs us of what directions leads to regions of higher probability. Note that it can also be computed without knowledge of $Z$. In HMC we view $U(\bm{x})$ as being a potential energy \citep{Duane1987}, which enables us to rewrite the target density as 
\begin{equation*}
\pi_{\bm{x}}(\bm{x}) \; = \; \frac{1}{Z}\exp\big(-U(\bm{x})\big).
\end{equation*}
We then interpret regions of higher potential energy as regions of lower probability. The state space $\mathcal{X}$ plays the role of the configuration manifold on which the dynamics are defined (i.e., it corresponds to $\mathcal{M}$ in the previous section). We define Hamiltonian dynamics on $\mathcal M$ by introducing a kinetic energy $K(\bm{p})=\frac{1}{2}g^{-1}(\bm{p},\bm{p})=\frac{1}{2}\bm{p}^T\bm{G}^{-1}\bm{p}$, and thus a Hamiltonian $H(\bm x,\bm p)=K+U$. We view the $d \times d$ matrix $\bm{G}$ as a covariance matrix and assume the momentum variables have the multivariate Gaussian density
\begin{equation*}
\pi_{\bm{p}}(\bm{p}) 
\; = \;
\mathcal{N}(\bm{p};\bm{0},\bm{G})
\; = \; 
\big((2\pi)^d|\bm{G}|\big)^{-\frac{1}{2}}\exp \big(-K(\bm{p})\big),
\end{equation*}
where $|\bm{G}|$ denotes the determinant of $\bm{G}$. The choice of the matrix $\bm{G}$ is critical for the performance of the algorithm, yet there is no general principle guiding its tuning. As a result it is often set to be the identity matrix. In section \ref{sec:RMHMC} we will see how the local structure of the target density may be used to choose a position-dependent $\bm{G}$. Define a joint density by 
\begin{equation*}
\pi(\bm{x},\bm{p}) \; = \; Z^{-1}\big{(}(2\pi)^d\mid \bm{G}\mid\big{)}^{-\frac{1}{2}}\exp\big(-H(\bm{x},\bm{p})\big)
\; = \;
\pi_{\bm{x}}(\bm{x})\pi_{\bm{p}}(\bm{p})
\end{equation*}
The HMC algorithm generates samples from this joint density.  Since the total energy $H$ is preserved along the flow, the joint probability $\pi(\bm{x},\bm{p})$ is constant along Hamiltonian trajectories. Here the Hamiltonian splitting $H=K+U$ is clearly applicable and we can hence use the leapfrog integrator. In practice, the simulation will not be exact since the leapfrog integrator is only approximately energy preserving, and a Metropolis step will be necessary to ensure we sample from the correct joint density. Given a current phase $(\bm{x}^k,\bm{p}^k) \in T^{\ast}\mathcal{X}$, the algorithm at iteration $k$ is given by:
\begin{description}
\item [1. ] Draw a momentum variable $\bm{p}^{k'}$ using $\pi_{\bm p}(\bm{p})$ i.e. $\bm{p}^{k'} \sim \mathcal{N}(\bm{0},\bm{G})$.
\item [2a.] Simulate dynamics with initial phase $(\bm{x}^k,\bm{p}^{k'})$ using the leapfrog integrator with fixed step-size $\tau$ for $L$ leapfrog steps, and flip the momentum of the resulting phase. This yields a proposal phase $(\bm{x}^{\ast},\bm{p}^{\ast})$.

\item [2b.] Accept the phase $(\bm{x}^{\ast},\bm{p}^{\ast})$ using a Metropolis step with probability
\begin{equation*}
\min \Big [1,\exp\Big(-H\big(\bm{x}^{\ast},\bm{p}^{\ast}\big)+H \big(\bm{x}^k,\bm{p}^{k'} \big)\Big) \Big ],
\end{equation*}
else keep the current phase: $\big(\bm{x}^{k+1},\bm{p}^{k+1}\big) = \big(\bm{x}^k,\bm{p}^{k'}\big)$. 
\end{description}

This algorithm simulates a Markov chain
\begin{marginnote}[]
\textbf{HMC steps:}\\
Step 1 of the algorithm is a momentum heatbath (Gibbs sampler). Step 2 is a Molecular Dynamics (MD) step (2a.) followed by a Markov Chain Monte Carlo (MC) rejection step (2b). Note this is sometimes called the Metropolis --Hastings step, although neither of them had much to do with it!
\end{marginnote}
 which if ergodic converges to the unique stationary density $\pi(\bm{x},\bm{p})$. The Markov chain can be shown to be geometrically ergodic under regularity assumptions \citep{Livingstone2015}. As $\pi_{\bm x}(\bm{x})$ is the marginal of our target density $\pi(\bm{x},\bm{p})$, we can then simply discard the auxiliary momentum samples to obtain samples of $\pi_{\bm x}(\bm{x})$.

Two parameters need to be tuned in order to apply HMC: the time-step $\tau$ and the trajectory length $L$. This tuning is often performed by running a few preliminary runs. Here, small time-steps will waste computational resources and slow down the exploration of the sample space, while large values of $\tau$ can lead to bad approximations of the trajectory which will dramatically reduce the acceptance probability. On the other hand, $L$ needs to be large enough to permit efficient explorations that avoid random walks and generate distant proposals; however too long trajectories may contain points in which the momentum sign flips, which can lead to poor exploration (think of a pendulum) \citep{Neal2011}. Several approach to tuning have been proposed in the literature, the most popular of which appears in \cite{Beskos2013} which proposes to tune parameter to maximise the computational efficiency as $d \rightarrow \infty$. Other approaches include the No U-Turn Sampler (NUTS) algorithm \citep{Hoffman2014},currently in use in the STAN programming software, and the use of Bayesian optimization \citep{Wang2013}. Finally, the shadow HMC algorithm, introduced in \S \ref{sec:Shadow_HMC}, has also been used to this effect \citep{Kennedy2012}.

\subsection{Relations to Stochastic Differential equations}

It is interesting to note that more information about the dynamics can be preserved (thus making the trajectory  more physical) if the full momentum resampling (the first step of HMC) is replaced by a partial momentum replacement \citep{Horowitz1991,Campos2015}. This enables us to sample more often as the trajectory length may be reduced to a single time-step without performing a random walk. Let $\bm{\xi} \sim \mathcal N (\bm{0},\bm{G})$ be Gaussian noise, the \textbf{generalised Hamiltonian Monte Carlo} (GHMC) algorithm is given by the following steps at each iteration $k$:
\begin{description}

\item [1. ] Rotate $(\bm p^k,\bm \xi^k)$ by an angle $\phi$.
\vspace{1mm}

\item [2a.] Perform MDMC step(s) to reach phase $(\bm x^*,\bm p^*)$.
\vspace{1mm}

\item [2b.] Flip the momentum, $F:\bm p^*\mapsto -\bm p^*$.
\vspace{1mm}

\item [2c.] Apply Metropolis accept/reject step.
\vspace{1mm}

\item [3. ] Flip the momentum, $F: \bm p^{k+1}\mapsto -\bm p^{k+1}$.

\end{description}

When $\phi =\pi/2$ we recover HMC. The first momentum flip is required to satisfy detailed balance. It however means that momentum is reversed in case of rejection which slows down the exploration if the rejection probability is non-negligible.

We now briefly mention links between HMC and algorithms based on stochastic differential equations (SDEs). If we consider the HMC algorithm in the special case of a single step of leapfrog integrator (i.e., $L=1$) with $K(\bm p)=\frac{1}{2}\bm p^T \bm p$, and drop the acceptance step, then each iteration $k$ is equivalent to: $$ \bm x^{k+1} =\bm x^{k}+ \tau \bm p^{k+1}- \frac{1}{2} \frac{ \partial U}{\partial \bm x}\tau ^2.$$ Defining $\varepsilon$ to be the square of the step size $\tau$ and the initial momentum to be Gaussian noise $\bm \xi$, we end up with a discretisation of the overdamped Langevin equation: $\bm x(\sqrt{\varepsilon}) =\bm x(0)+ \sqrt{\varepsilon} \bm \xi-  \frac{1}{2} \frac{ \partial U}{\partial \bm x}\varepsilon.$ If we add the Metropolis-Hastings step, this algorithm corresponds to the MALA algorithm previously discussed, which is an exact version of the Langevin algorithm (in the sense that there is no discretisation error). 

Hamiltonian Monte Carlo can also be related to higher order SDE: consider the following second order Langevin dynamics defined on a Riemannian manifold (with diffusion defined by a vector field $v$), $$d \bm x \; = \; \bm vdt, \quad \quad d\bm v \;=\; -\gamma(\bm x,\bm v)dt-\bm G^{-1}(\bm x)\frac{\partial U}{\partial \bm x}dt-\bm vdt + \sqrt{2\bm G^{-1}(\bm x)}dW.$$ 
\begin{marginnote}[]
\textbf{Christoffel Symbols:} \\
{The Christoffel symbols are defined by 
$\Gamma^{k}_{ij} = \sum_{r} \frac{1}{2}g^{kr}\\(\partial_j g_{ir}+ \partial_ig_{rj} 
-\partial_rg_{ij})$.
They give information about the curvature of the manifold.
}\end{marginnote}	
Here, $W$ is a standard Wiener process and $d\bm v+\gamma dt$ is the covariant time derivative (physically, the acceleration) of the velocity and thus $\gamma$ has $k^{\text{th}}$ component $\sum_{ij}\Gamma^k_{ij}v^iv^j$ where $\Gamma^{k}_{ij}$ are the Christoffel symbols. The SDE may be transferred to phase space using $ \bm p = G(\bm x) \bm v$. The invariant distribution of this diffusion  may be easily shown to be $\pi(\bm x, \bm p) \propto |G(\bm x)|^{-1} \exp \big(-\frac{1}{2}\bm p^T \bm G^{-1}(\bm x)\bm p-U(\bm x)\big)$. Thus setting $U(\bm x)=-\log \pi_{\bm x}(\bm x)-\frac 1 2 \log | \bm G(\bm x)|$ gives $\pi_{\bm x}(\bm x)$ as the marginal distribution. 

To simulate from the SDEs above, it is convenient to use schemes that rely on Lie--Trotter splitting \citep{Abdulle2015}, in which the numerical method will be of the form $\Phi_{\tau} \circ \Psi_{\tau}$, where $\Phi_{\tau}$ is an integrator for the deterministic part and $\Psi_{\tau}$ 
for the stochastic part. Notice the stochastic part is a conditioned Ornstein--Uhlenbeck process and corresponds to partial momentum refreshment. We can then use a symplectic integrator to sample from $\pi(\bm{x},\bm{p})$ via RMHMC, which we introduce below. 
\begin{marginnote}[]
\textbf{SDEs on Manifolds:} \\
{The SDE represents the acceleration of a particle on a manifold under the influence of a noisy potential and subject to a friction term $v\,dt$. } \end{marginnote}


\section{RIEMANNIAN MANIFOLD HMC}\label{sec:RMHMC}

We have seen how HMC uses gradient information from the target density to improve the exploration of the state space. Recently, \cite{Girolami2011} introduced a method, called \textbf{Riemmanian Manifold Hamiltonian Monte Carlo} (RMHMC), that uses the higher order information available so that the transition density adapts to the local geometry of the target density (see also \citep{Livingstone2014}). A notion of distance is defined between points in state space, so that smaller steps are performed in directions in which the target density changes rapidly. This method hence borrows tools from the field of information geometry \citep{Amari1987}.

In the original version of this algorithm \cite{Girolami2011} considered sampling from a Bayesian posterior density: after observations $\bm{y}=(y_1,\ldots,y_n)$ have been made, the target density $\pi_{\bm x}(\cdot)$ may be updated to a posterior $\pi(\cdot | \bm{y})$ through a likelihood function $\mathcal{L}(\cdot | \bm{x})$ by the means of Bayes theorem, $\pi(\bm{x}| \bm{y}) \propto \mathcal{L}(\bm{y}|\bm{x}) \pi_{\bm x}(\bm{x})$. 
They took advantage of the fact that the likelihood function defines a statistical model with parameters $\bm{x}$, that is for each $\bm{x}$, $\mathcal{L}(\cdot|\bm{x})$ is a density. Under mild conditions \citep{Amari1987} the statistical model $\mathcal{S}:=\{\mathcal{L}(\cdot|\bm{x}): \bm x \in \mathcal{X}\}$ is a manifold with global coordinates $\bm x$. Hence each point on the original manifold $\mathcal{X}$ is now associated to the density $\mathcal{L}(\cdot|\bm{x})$. 

On the statistical manifold $\mathcal{S}$, it is common to identify a vector field $v= \sum_{j=1}^d v^j\partial_{j}$ with the random variable $v^{(1)}= \sum_{j=1}^d v^j \frac{\partial l(\cdot|\bm x)}{\partial x^j}$, where $l(\cdot| \bm x):=\log \mathcal{L}(\cdot|\bm{x})$ is the log-likelihood. This is called the $1$-representation of the tangent space. We can define a natural inner product on the tangent spaces of $\mathcal{S}$ called the \textbf{Fisher metric}, by defining an inner product on the corresponding $1$-representations: $g_{\bm{x}}(u,v):=\mathbb E_{l(\cdot|\bm{x})} [ u^{(1)}v^{(1)}]$. As a result, the configuration manifold $\mathcal{X}$ acquires the Riemannian metric $g$ and thus a natural concept of distance between densities associated to $\bm{x} \in \mathcal{X}$.

To tailor the metric to Bayesian problems, which are common in MCMC, \cite{Girolami2011} proposed a variant of the Fisher metric which adds the negative Hessian of the log-prior: $$\bm G_{ij}(\bm x) \; = \; \bm F_{ij}(\bm x)-\frac{\partial^2}{\partial x^i \partial x^j} \log \pi_0(\bm x),$$
where $\pi_0(\bm x)$ is the prior density and $\bm F$ the Fisher metric. The kinetic energy is defined using this metric $\bm{G}(\bm{x})$, so that the momentum variable is now Gaussian with a position-dependent covariance matrix $\pi(\bm{p}| \bm{x})= \mathcal{N}\big{(}\bm{0},\bm{G}(\bm{x})\big{)}$ which can help mitigate some of the scaling and tuning issues associated to HMC.  The Hamiltonian on the Riemannian manifold is $$H(\bm{x},\bm{p}) \; = \; U(\bm{x})+\frac{1}{2}\log \big{(}(2\pi)^d | \bm{G}(\bm{x})|\big{)}+\frac{1}{2}\bm{p}^T\bm{G}^{-1}(\bm{x})\bm{p}.$$
The joint density $\pi(\bm{x},\bm{p}):= \exp{\big(-H(\bm{x},\bm{p})\big)}=\pi_{\bm x}(\bm{x})\pi(\bm{p} | \bm{x})$ still has the desired target $\pi_{\bm x}(\bm{x})$ as marginal density, but the Hamiltonian is no longer separable. Thus, the leapfrog integrator is no longer symplectic and reversible; instead we use a \textbf{generalised leapfrog algorithm}. At each iteration $k$, the algorithm is given by: 
\begin{eqnarray*}
\bm{p}^{k+\frac{1}{2}} 
& = &
 \bm{p}^k-\frac \tau 2 \frac{\partial H}{\partial \bm{x}}\big(\bm{x}^k,\bm{p}^{k+\frac{1}{2}}\big), \\[0.5em]
\bm{x}^{k+1} 
& = &
\bm{x}^k+\frac\tau2 \Big{(}\frac{\partial H}{\partial \bm{p}}\big(\bm{x}^k,\bm{p}^{k+\frac{1}{2}}\big)+\frac{\partial H}{\partial \bm{p}}\big(\bm{x}^{k+1},\bm{p}^{k+\frac{1}{2}}\big)\Big), \\[0.5em]
\bm{p}^{k+1} 
& = &
\bm{p}^{k+\frac{1}{2}}-\frac\tau2  \frac{\partial H}{\partial \bm{x}}\big(\bm{x}^{k+1},\bm{p}^{k+\frac{1}{2}}\big) . 
\end{eqnarray*}
As these equation are implicit we must resort to fixed point iterations method. Notice that the additional information provided by the local geometry of the statistical manifold can lower the correlation between samples and increase the acceptance rate. Such an advantage will be particularly useful in high dimensions, where concentration of measure makes sampling very challenging (even though the computational cost of RMHMC will also increase as $O(d^3)$).   

Recent advances have included replacing the underlying Lebesgue measure by the Hausdorff measure; as a result $H$ becomes $H(\bm{x},\bm{p})=U_{\mathcal H}(\bm x)+\frac{1}{2}\bm{p}^T\bm{G}^{-1}(\bm{x})\bm{p}$, where $U_{\mathcal H}(\bm x) := -\log \pi_{\mathcal H}(\bm x)$ is the potential energy of the target density $\pi_{\mathcal H}$ with respect to the Hausdorff measure. 
This has the advantage that the method of splitting Hamiltonian can then be used to construct a geodesic integrator \citep{Byrne2013}. The use of Lagrangian dynamics \citep{Fang2014,Lan2015} has also been proposed to obtain integrators which are not volume preserving but which have lower computational costs.
Finally, other Riemannian metrics that do not rely on a Bayesian setting have been studied, often with the aim of improving the sampling of multi-modal densities \citep{Lan2014,Nishimura2016Tempered}.


\section{SHADOW HAMILTONIANS} \label{sec:Shadow_HMC}

We now discuss a remarkable property of symplectic integrators: the existence of a shadow Hamiltonian that is exactly conserved by the symplectic integrator (in the sense of an asymptotic expansion). The study of this quantity was inspired by backward-error analysis of differential equations and is used in molecular dynamics but is mostly unknown in the statistics literature. This is partly due to the geometric notions required to define it, in particular the Poisson bracket and its Lie algebra structure. In this section, we provide an intuitive introduction to the shadow Hamiltonian. This is complemented by a section in the supplementary material, in which we define those advanced notions more carefully.

\begin{figure}
\includegraphics[width=0.4\textwidth,clip,trim = 0 0 0 0]{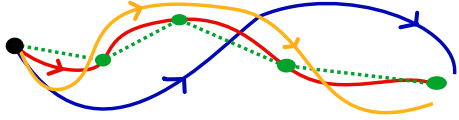}
\caption{Shadow Hamiltonian Monte Carlo - The blue line gives the exact trajectory. The dotted green line is the numerical method and exactly follows the shadow trajectory (in red). The orange line is the approximate shadow trajectory.}
\label{fig:shadow_Hamiltonian}
\end{figure}

We have seen that the leapfrog integrator (and other symplectic integrators) do not exactly preserve the Hamiltonian $H$. Over non-infinitesimal times this causes the simulated trajectory to diverge from the exact Hamiltonian trajectory.
We may expect the energy along the approximate trajectory to diverge linearly with the trajectory length (number of steps). However, in practice, the energy does not diverge but merely oscillates around the correct energy even for very long trajectory lengths. The reason for this is that there is a nearby Hamiltonian that is exactly conserved by the discrete integrator. 
That is we can find a \textbf{shadow Hamiltonian} $\tilde{H}_{\tau}$ that is constant along the simulated trajectory (see Fig. \ref{fig:shadow_Hamiltonian}). The shadow Hamiltonian is defined as an asymptotic expansion which is exponentially accurate (for small enough step-size). The aim of \textbf{Shadow Hamiltonian Monte Carlo} (SHMC) \citep{Izaguirre2004} is to sample from a distribution with density close to $e^{-\tilde H_{\tau}}$ in order to improve the acceptance rate, and then correct for the fact we are not sampling from the desired target density by re-weighting.

Using the Baker--Campbell--Hausdorff formula it is possible to build Hamiltonians that are arbitrarily close to the shadow Hamiltonian \citep{Skeel2001}, i.e. they satisfy $\tilde H_{[2d]} =  \tilde{H}_{\tau}+O(\tau^{2d})$ (the square bracket notation indicates the order of the approximation). A difficulty is that the shadow Hamiltonian is not a sum of a kinetic and a potential term and therefore the momentum refreshment step is no longer just sampling from a Gaussian distribution. The SHMC algorithm instead samples from the new target density $
\rho_M(\bm{x},\bm{p}) := (1/Z_{\rho})e^{-H_M(\bm{x},\bm{p})}$, defined by the Hamiltonian $H_M(\bm{x},\bm{p}) := \max \big\{H(\bm{x},\bm{p}),\tilde H_{[2d]}(\bm x,\bm p)-a \big\}$ where $a$ is a constant parameter that bounds the allowed difference between $\tilde{H}_{\tau}$ and $\tilde{H}(\bm{x},\bm{p})$ and needs to be tuned. 
The purpose of introducing this maximum is that it is bounded below by $H$. We can therefore generate Gaussian samples from $H$ and then use rejection sampling (also called von Neuman's rejection method \citep{Robert2004}) to convert these into samples from $\rho_M$. When $a$ is large and positive $H_M$ is essentially the same as $H$ and we will achieve a high acceptance rate for the rejection sampler, while when $a$ is large and negative we will approximate well the shadow but will have a low acceptance rate. Hence the tuning of $a$ is critical. Given a current state $(\bm{x},\bm{p})\in T^* \mathcal{X}$, the algorithm proceeds as follows:	
\begin{enumerate}
\item Draw a new momentum $\bm{p}'$ from $\mathcal{N}(\bm{0},\bm{I})$ and accept it with probability 
\begin{equation*}
\frac{e^{-H_M(\bm x,\bm p')}}{e^{-H(\bm x,\bm p')}} =\min \Big[1, \frac{\exp \big(a-\tilde H_{[2d]}(\bm{x},\bm{p}')\big)}{\exp \big( -H(\bm x,\bm p')\big)} \Big]
\end{equation*}
Repeat until a $\bm{p}'$ is accepted. This is simply rejection sampling.  
\vspace{1mm}
\item Simulate Hamiltonian mechanics with initial phase $(\bm{x},\bm{p}')\in T^*\mathcal{M}$ and Hamiltonian $H$ using a symplectic time-reversible integrator. This yields a proposed configuration $(\bm{x}^{\ast},\bm{p}^{\ast})$, which we accept with probability $\min \left\{1,\frac{\rho(\bm{x}^{\ast},\bm{p}^{\ast})}{\rho(\bm{x},\bm{p}')}\right\}$, else keep the old phase $(\bm x,\bm p)$.
\end{enumerate}

\begin{marginnote}
\textbf{SHMC steps:}\\
As before, step $1$ is a momentum heatbath (Gibbs sampler) and step $2$ is a MD-MC step.
\end{marginnote}To calculate the sample average, a re-weighting process is necessary to compensate for the fact we are sampling from the wrong distribution. To do this we re-weight the samples generated with a factor $c_k := \exp\big(\tilde H (\bm x^l,\bm p^l)-H (\bm x^l,\bm p^l)\big)$. The main advantage of this method is that the Metropolis acceptance rate will be much closer to one. On the other hand the momentum refreshment step can become expensive and the variance of the sample average will be large if the factors $c_k$ are not close to one.

There are however several issues surrounding SHMC. While the acceptance rate is greatly improved in the Metropolis step, SHMC samples from distributions with non-separable Hamiltonians which makes the momentum sampling more expensive. 
Moreover it introduces a new parameter $a$ to balance the acceptance rates of the two steps. \citep{Sweet2009} built a variant in which a canonical transformation (symplectomorphism) is used to change coordinates in order to get a separable Hamiltonian. Alternatively, a \textbf{Generalised Shadow Hamiltonian Monte Carlo} algorithm has also been proposed\citep{Akhmatskaya2008}.

Finally, we note that the shadow Hamiltonian can be used to tune the parameters of HMC \citep{Kennedy2012}. The variance $\text{Var}(H-\tilde H)$ may be expressed as a function of Poisson brackets and integrator parameters, and it turns out that for extensive systems the Poisson brackets are almost constant. It follows that we can  tune the parameters of complicated symmetric symplectic integrators and minimise this variance by simply measuring the appropriate Poisson brackets. 
\section{RECENT RESEARCH DIRECTIONS} \label{sec:recent_advances}

In this final section, we discuss some of the most recent research directions in HMC including stochastic gradient methods and algorithms in infinite dimensions. These have been developed to deal with the increasing size of datasets and the complexity of models that scientists have to deal with.

\subsection{Stochastic Gradient Markov Chain Monte Carlo}

One of the major issues in the use of MCMC methods in a Bayesian context is the size of datasets. Imagine that we have i.i.d. observations $\bm{y}=(\bm{y}_1,\ldots,\bm{y}_n)$ and are interested in the posterior density over some parameter $\bm{x} \in \mathcal{X} \subseteq \mathbb{R}^d$ (here, we assume we have pre-defined some prior $\pi_0$): $
\pi(\bm{x}|\bm{y}) \; \propto \; \pi_0(\bm{x}) \prod_{j=1}^n \mathcal{L}(\bm{y}_j|\bm{x})$. 
Clearly, if $n$ is very large then the posterior $\pi(\bm{x}|\bm{y})$ and the score functions $\partial_{i} \log \pi(\bm{x}|\bm{y})$ will be computationally expensive to evaluate, rendering MCMC costly. To tackle this issue, \cite{Welling2011} suggested making use of small subsets of the entire dataset (called mini-batches) to compute the score functions, making this inference tractable once again. Although this methodology was originally developed for MALA, it was later extended to HMC algorithms \citep{Chen2014,Chen2015,Ma2015}. It is however important to note that these algorithms are not exact (in the sense that they only target an approximate target density), and the the bias could be large and very difficult to assess a-priori \citep{Teh2014,Vollmer2015,Betancourt2015Subsampling}.

\subsection{Infinite-Dimensional HMC}

Recently \citep{Beskos2011,Cotter2013} proposed to deal with the degrading performance of HMC in very high dimensions by building a HMC algorithm which samples from a measure $\mu$ on an infinite-dimensional Hilbert space, such that our target measure $\pi$ is a finite-dimensional projection of $\mu$. Informally, we can think of infinite-dimensional probability distributions as being a distribution on functions (e.g. Gaussian processes or Dirichlet processes). Measures of this form appear in a wide range of applications, from fluid dynamics to computational tomography; and more generally in Bayesian inverse problems \citep{Stuart2010,Beskos2016}.

The algorithm samples from a measure $\mu$ on a separable Hilbert space $\mathcal H$, which is defined by its Radon--Nikodym derivative with respect to a dominating Gaussian measure $\mu_0$, as given by $\frac{d \mu}{d\mu_0}(x) \propto \exp \big (-\Phi(x)\big )$ for some potential function $\Phi : \mathcal H \rightarrow \mathbb R$. Looking at HMC this way removes the dependence on the dimension $d$ of the projection, as the algorithm is defined directly over an infinite-dimensional space. Specifically, it allows for efficient sampling from target measures in very large dimensions and the acceptance rate does not tend to $0$ as $d\rightarrow \infty$, since the algorithm is well-defined in that limit.

\section{CONCLUSION}

The use of Differential Geometry in Statistical Science dates back to the early work of C.~R.~ Rao in the 1940's when he sought to assess the natural distance between population distributions \citep{Rao45statmanifold}. 
The Fisher--Rao metric tensor defined the Riemannian manifold structure of probability measures and from this local manifold geodesic distances between measures could be properly defined. This early work was then taken up by many authors within the statistical sciences with an emphasis on the study of the efficiency of statistical estimators \cite{Efron1982,Barndorff-Nielsen1986,Amari1987,Critchley1993,Murray1993}. The area of Information Geometry \citep{Amari1987} has developed substantially and has had major impact in areas of applied statistics such as Machine Learning and Statistical Signal Processing. 

In 2010, a landmark paper was published where Langevin diffusions and Hamiltonian dynamics on the manifold of probability measures were defined to obtain Markov transition kernels for Monte Carlo based inference \citep{Girolami2011}. This work was motivated by the many challenges presented by contemporary problems of statistical inference, such as for example inference over partial differential equations describing complex physical engineering systems. 

This review has aimed to provide an accessible introduction to the necessary differential geometry, with a specific focus on the elements required to formally describe Hamiltonian Monte Carlo in particular. This formal understanding is necessary to gain insights into more advanced methods, including Shadow Hamiltonian and Riemann Manifold Hamiltonian methods. This should also be of interest to readers interested in the development of new methods that seek to address the growing list of challenges modern day statistical science is being called upon to address. More generally, we believe the use of geometry is essential to even attempt to tackle sampling issues related to the curse of dimensionality and concentration of measure in for example Deep Learning.


\section*{ACKNOWLEDGMENTS}
The authors are grateful to the Alan Turing Institute for supporting the development of this work. AB was supported by a Roth scholarship from the Department of Mathematics at Imperial College London. FXB was supported by the EPSRC grant [EP/L016710/1]. MG was supported by the EPSRC grants [EP/J016934/3, EP/K034154/1, EP/P020720/1], an EPSRC Established Career Fellowship, the EU grant [EU/259348], a Royal Society Wolfson Research Merit Award and the Lloyds Register Foundation Programme on Data-Centric Engineering, ADK was supported by STFC Consolidated Grant ST/J000329/1.

{ \scriptsize
\bibliographystyle{plainnat}
\bibliography{HMC_bib}
}

\newpage

\section{ONLINE SUPPLEMENT} \label{sec:Appendices}

The following sections complement the theoretical background on differential geometry and Hamiltonian dynamics as defined in \S\ref{sec:geometry_physics} and \S\ref{sec:Hamiltonian_Monte_Carlo}. This material is, for example, required to define the shadow Hamiltonians which were described in \S\ref{sec:Shadow_HMC}. However, is not necessary to understand the main paper, and only provides additional details for the interested reader. 

This appendix is structured as follows. The first part generalises the concept of differential forms in \S \ref{appendix:diff_forms_ext_derivatives} and uses it to provide a more general definition of Hamiltonian dynamics in \S \ref{appendix:Hamiltonian_field_Poisson_brackets}. Then, \S\ref{appendix:SHMC} and \S \ref{appendix:practical_integrators} formally discuss shadow Hamiltonians and discusses the construction of numerical integrators. Finally, we conclude with a discussion of Hamiltonian dynamics on Lie groups in \S\ref{appendix:Hamiltonia_mechanics_Lie_groups}
and Lagrangian dynamics: \S\ref{appendix:Lagrangian_mechanics}, and how these may be of interest in an MCMC context.

\subsection{Differential Forms and Exterior Derivatives} \label{appendix:diff_forms_ext_derivatives}

We have already defined the simplest cases of differential forms in \S \ref{sec:manifolds_differential_forms},
including $1$-forms and $2$-forms. Let $\mathcal X(\mathcal M)$ denote the space of vector fields on $\mathcal M$. A differential \textbf{$\bm{k}$-form} $\omega^k$ is a map $\omega^k : \mathcal X(\mathcal M)^k \rightarrow C^{\infty}(\mathcal M)$ which is multilinear (i.e., linear with respect to $C^{\infty}(\mathcal M)$ in each argument) and antisymmetric, that is it changes sign whenever two arguments are exchanged.

We denote by $\Omega^k(\mathcal{M})$ the space of differential $k$-forms. Given $\alpha\in\Omega^k$ and $\beta\in\Omega^s$ we can define the \textbf{tensor product} to be the multilinear map $\alpha \otimes \beta: \mathcal X( \mathcal M)^{k+s}\rightarrow C^{\infty}(\mathcal M)$ by 
$$ \alpha \otimes \beta(v_1,\ldots,v_{k+s}) \; := \; \alpha(v_1,\ldots,v_k)\beta(v_{k+1},\ldots,v_{k+s}).$$ 

The \textbf{wedge product} $\wedge : \Omega(\mathcal M)^k \times \Omega(\mathcal M)^s \rightarrow \Omega(\mathcal M)^{k +s}$ of $\alpha$ and $\beta$ is defined to be the antisymmetrisation:
$$ \alpha \wedge \beta(v_1,\ldots,v_{k+s}) \; := \; \frac{1}{k! s!} \sum_{\sigma \in S_{k+s}} (-1)^{\sigma}\alpha(v_{\sigma(1)},\ldots,v_{\sigma(k)})\beta(v_{\sigma(k+1)},\ldots,v_{\sigma(k+s)}), $$
 where $S_n$ is the group of permutations of $1,\ldots,n$ and $(-1)^{\sigma}$ denotes the sign of the permutation $\sigma$.
 
The normalization is required to ensure that the wedge product is associative, since 

\begin{equation*} 
\begin{split}
 \Bigl(&(\alpha\wedge\beta)\wedge\gamma\Bigr)(v_1,\ldots,v_{k+s+t})\\
    &= {1\over(k+s)!t!} \sum_{\tau\in S_{k+s+t}} (-1)^\tau
      (\alpha\wedge\beta)(v_{\tau_1},\ldots,v_{\tau_{k+s}})
      \gamma(v_{\tau_{k+s+1}},\ldots,v_{\tau_{k+s+t}}) \\
    &= {1\over(k+s)!t!} \sum_{\tau\in S_{k+s+t}} (-1)^\tau {1\over k!s!}\\
      &\sum_{\sigma\in S_{k+s}} (-1)^\sigma
        \alpha(v_{\tau_{\sigma_1}},\ldots,v_{\tau_{\sigma_k}})
        \beta(v_{\tau_{\sigma_{k+1}}},\ldots,v_{\tau_{\sigma_{k+s}}})
        \gamma(v_{\tau_{k+s+1}},\ldots,v_{\tau_{k+s+t}}) \\
    &= {1\over(k+s)!k!s!t!} \sum_{\sigma\in S_{k+s}}\\ &\sum_{\tau\in S_{k+s+t}}
        (-1)^{\tau'} \alpha(v_{\tau'_1},\ldots,v_{\tau'_k})
        \beta(v_{\tau'_{k+1}},\ldots,v_{\tau'_{k+s}})
        \gamma(v_{\tau'_{k+s+1}},\ldots,v_{\tau'_{k+s+t}}),
\end{split}
\end{equation*}

where we choose $\tau'\in S_{k+s+t}$ such that
$$ \tau'_i= \tau_{\sigma_i}  \; \text{for} \; 1\leq i\leq k+s,       \tau'_i= \tau_i  \; \text{for} \; k+s+1\leq i\leq k+s+t,$$
so
$$ \sum_{\sigma\in S_{k+s}} \sum_{\tau\in S_{k+s+t}} f(\tau')
  = (k+s)! \sum_{\tau'\in S_{k+s+t}} f(\tau')$$
and $(-1)^\sigma (-1)^\tau = (-1)^{\tau'}$.  Hence

\begin{equation*} 
\begin{split}
\Bigl(&(\alpha\wedge\beta)\wedge\gamma\Bigr)(v_1,\ldots,v_{k+s+t}) \\
    &= {1\over k!s!t!} \sum_{\tau'\in S_{k+s+t}} (-1)^{\tau'}
        \alpha(v_{\tau'_1},\ldots,v_{\tau'_k})
        \beta(v_{\tau'_{k+1}},\ldots,v_{\tau'_{k+s}})
        \gamma(v_{\tau'_{k+s+1}},\ldots,v_{\tau'_{k+s+t}}) \\
    &= {1\over k!s!t!(s+t)!} \sum_{\sigma'\in S_{s+t}}\\ & \sum_{\tau''\in S_{k+s+t}}
        (-1)^{\tau'} \alpha(v_{\tau''_1},\ldots,v_{\tau''_k})
        \beta(v_{\tau''_{\sigma'_1}},\ldots,v_{\tau''_{\sigma'_s}})
        \gamma(v_{\tau''_{\sigma'_{s+1}}},\ldots,v_{\tau''_{\sigma'_{s+t}}}),
\end{split}
\end{equation*}

where we choose $\tau''\in S_{k+s+t}$ such that
$$ \tau''_i = \tau'_i  \; \text{for} \; 1\leq i\leq k, 
                  \tau''_i =  \tau'_{\sigma'^{-1}_{i-k}+k}  \; \text{for} \; k+1\leq i\leq k+s+t,$$
whence
$$  \tau'_i = \tau''_i  \; \text{for} \; 1\leq i\leq k, 
                   \tau'_i = \tau''_{\sigma'_{i-k}+k}  \; \text{for} \; k+1\leq i\leq k+s+t,
$$
so
$$
  \sum_{\sigma'\in S_{s+t}} \sum_{\tau''\in S_{k+s+t}} f(\tau')
  = (s+t)! \sum_{\tau'\in S_{k+s+t}} f(\tau')
$$
and $(-1)^{\tau''} (-1)^{\sigma'} = (-1)^{\tau'}$. Therefore

\begin{equation*} 
\begin{split}
\Bigl(&(\alpha\wedge\beta)\wedge\gamma\Bigr)(v_1,\ldots,v_{k+s+t}) \\
    &= {1\over k!(s+t)!} \sum_{\tau''\in S_{k+s+t}} (-1)^{\tau''}
      \alpha(v_{\tau''_1},\ldots,v_{\tau''_k})
      {1\over s!t!} \sum_{\sigma'\in S_{s+t}} (-1)^{\sigma'}
        \beta(v_{\tau''_{\sigma'_1}},\ldots,v_{\tau''_{\sigma'_s}})
        \gamma(v_{\tau''_{\sigma'_{s+1}}},\ldots,v_{\tau''_{\sigma'_{s+t}}}) \\
    &= {1\over k!(s+t)!} \sum_{\tau''\in S_{k+s+t}} (-1)^{\tau''}
         \alpha(v_{\tau''_1},\ldots,v_{\tau''_k})
         (\beta\wedge\gamma)(v_{\tau''_{k+1}},\ldots,v_{\tau''_{k+s+t}}) \\
    &= \Bigl(\alpha\wedge(\beta\wedge\gamma)\Bigr)(v_1,\ldots,v_{k+s+t}).
\end{split}
\end{equation*}

The \textbf{exterior derivative} $d_k: \Omega^k(\mathcal{M}) \rightarrow \Omega^{k+1}(\mathcal{M})$ is defined by 
\begin{eqnarray*}
d_k\omega(v_1,\ldots,v_{k+1}) 
& := & \sum_j (-1)^jv_j\big (\omega (v_1,\ldots,\hat v_j,\ldots,v_{k+1}) \big) \\
& &  +\sum_{j < k} (-1)^{k+j}\omega \big([v_j,v_k],v_1,\ldots,\hat v_j,\ldots,\hat v_k,\ldots,v_{k+1} \big),
\end{eqnarray*}
where the hat indicates we omit the corresponding entry, and $[v_j,v_k]$ is the vector field commutator defined by $[v_j,v_k]f:=v_j\big(v_k(f)\big)-v_k\big(v_j(f)\big)$.
Vector fields are linear differential operators so their commutators are also vector fields since all the second derivative operators cancel. We will henceforth omit the subscript $k$ on the exterior derivative operator. The exterior derivative has the following properties:
 
\begin{enumerate}
\item If $f\in C^{\infty}(\mathcal{M})=\Omega^0(\mathcal M)$, $df \in \Omega^1(\mathcal{M})$ is the differential $df(v):=v(f)$.

\item If $\alpha,\beta \in \Omega(\mathcal{M})$ then  $d(\alpha+\beta)=d\alpha +d\beta$. 

\item $d^2 :=d\circ d=0$.

\item If  $\alpha \in \Omega^k(\mathcal{M})$ and $\beta \in \Omega(\mathcal{M})$, then $d\big (\alpha \wedge \beta \big ) = (d\alpha ) \wedge \beta + (-1)^k\alpha \wedge d\beta$.
\end{enumerate}
An alternative definition of the exterior derivative is that it is the unique operator with these properties. The coordinate expression for the exterior derivative of a $1$-form $\alpha = \alpha_i dx^i$ and a 2-form $\omega=\omega_{ij}dx^i\wedge dx^j$ can be easily found from the properties above: $$d \alpha = (\partial_{j}\alpha_i)dx^j\wedge dx^i, \quad \quad \omega=(\partial_{k}\omega_{ij} )dx^k\wedge dx^i \wedge dx^j.$$
Differential $k$-forms are objects that may be integrated over a $k$-dimensional surface (a $k$-chain). Stokes' theorem relates the integral of a $k$-form $\omega$ over the $k$-dimensional boundary $\partial D$ of a $(k+1)$-dimensional surface $D$ to the integral of $d\omega $ over $D$: $$\int_D d \omega = \int_{\partial D} \omega.$$ 
Finally, we introduce the de Rham cohomology.
 
The exterior derivative is a homomorphism that maps $k$-forms to $(k+1)$ forms and thus defines a sequence 
$$ \Omega^0(\mathcal M) \;  \overset{d} \longrightarrow  \; \Omega^1({\mathcal M}) \; \overset{d} \longrightarrow \; \Omega^2(\mathcal M) \; \overset{d} \longrightarrow \; \cdots \; \overset{d} \longrightarrow  \; \Omega^k({\mathcal M}) \; \overset{d} \longrightarrow \; \cdots $$
Since $d^2 = d^{k+1}\circ d^k = 0$ for all $k$, this sequence is called a \textbf{cochain complex}, and we can define the $k^{th}$ \textbf{de Rham cohomology group} to be the quotient group $H^k(\mathcal M) =\mathop{\rm{Ker}}(d_{k+1})/\mathop{\rm{Im}}(d_k)$. A $k$-form $\omega$ is exact if there exists a $(k-1)$-form $\alpha$ such that $\omega=d\alpha$ and that it is closed if $d\omega=0$. 
Thus $\omega$ is exact iff $\omega \in \mathop{\rm{Im}}(d_k)$, and $\omega$ is closed iff $\omega \in \mathop{\rm{Ker}}(d_{k+1})$. Clearly all exact forms are closed (by property 3 above). We define an equivalence relation on the space of closed forms by identifying two closed forms if their difference is exact. The set of equivalence (cohomology) classes induced by this relation is precisely the de Rham cohomology group above.

\subsection{Hamiltonian Vector Fields and Poisson Brackets}\label{appendix:Hamiltonian_field_Poisson_brackets}

In the main text, we introduced a special case of Hamiltonian dynamics where the Hamiltonian was the sum of a potential and a kinetic energy, and the kinetic energy was defined by the Riemannian metric. We now introduce Hamiltonian mechanics more generally. 

Given a symplectic manifold $(\mathcal M,\omega)$ (so $\omega$ is a closed, non-degenerate differential 2-form) and any $0$-form $H\in C^{\infty}(\mathcal M)$, we can define a \textbf{Hamiltonian vector field} $\hat H:\mathcal{M} \rightarrow T \mathcal{M}$ by 
$$dH(\cdot)  \; = \; \omega(\hat H, \cdot).$$
where $\hat H(p) \in T_p\mathcal{M}$ for all $p \in \mathcal{M}$.

The triple $(\mathcal{M},\omega, H)$ is called a \textbf{Hamiltonian system}. Notice that Hamiltonian systems do not require any Riemannian metric to be defined (in particular in this general setting, momentum field are not necessarily related to velocity fields).

A \textbf{Lie bracket} on a vector space $V$, is a product $[\cdot,\cdot]:V\times V\rightarrow V$ which is antisymmetric, linear in each entry, and satisfies the Jacobi identity: $$\big[A,[B,C]\big]+\big[B,[C,A]\big]+\big[C,[A,B]\big]
\; = \;
0.$$
We call $\big(V,[\cdot,\cdot]\big)$ a \textbf{Lie algebra}. The \textbf{Poisson bracket} $\{\cdot,\cdot \}:C^{\infty}(\mathcal M)\times C^{\infty}(\mathcal M) \rightarrow C^{\infty}(\mathcal M)$ of two $0$-form is defined as 
$$\{ A, B \} \; := \; -\omega(\hat A, \hat B).$$
The vector space $C^{\infty}(\mathcal M)$ equipped with this product is a Lie algebra. The Jacobi identity is a consequence of the fact that the symplectic $2$-form is closed, $d\omega=0$. A key property of Hamiltonian mechanics is that the commutator of two Hamiltonian vector fields $\big[\hat A,\hat B \big]$, defined by $\big[\hat A,\hat B \big]f:=\hat A \big (\hat B (f)\big)-\hat B\big ( \hat A (f) \big )$, is itself a Hamiltonian vector field and that furthermore
$$ \widehat{\{A,B\}} = [\hat A,\hat B],$$
that is the Hamiltonian vector field of the Poisson bracket $\{A,B\}$ is the commutator of the Hamiltonian vector fields of $A$ and $B$.

\subsection{Lie Groups and Baker-Campbell-Hausdorff formula}

\subsubsection{Lie Derivative}
A \textbf{tensor field} is a $C^{\infty}(\mathcal M)$--multilinear map $T:\mathcal X(\mathcal M)\times \cdots \times \mathcal X(\mathcal M) \times \Omega^1(\mathcal M)\times \cdots \times \Omega^1(\mathcal M)\rightarrow C^{\infty}(\mathcal M)$. We also consider the special case of $0$-forms to be tensors. To define how tensor fields vary along the Hamiltonian trajectories, we define a differential operator called the \textbf{Lie derivative}. To do so we need the concepts of local flow and induced map.

 Let $U\in \mathcal M$ be an open set. A \textbf{local flow} is a smooth map $\sigma : (-\varepsilon,\varepsilon)\times U\rightarrow \mathcal M$, such that (i) for each $t\in (-\varepsilon,\varepsilon)$, $\sigma_t(\cdot):=\sigma(t,\cdot):U\rightarrow \mathcal M$ is a diffeomorphism (a smooth map with smooth inverse) onto its image, and (ii) $\sigma_t$ is a homomorphism in $t$.  If $v \in \mathcal X(\mathcal M)$, it is possible to prove that $v$ may be generated by a local flow around each point $p\in \mathcal M$, which means that $\exists U$ such that $\forall p\in U$, $v_p$ is equal to the tangent vector of the curve $t\mapsto \sigma_t(p)$ at $t=0$.

Given any map $f:\mathcal N \rightarrow \mathcal S$ between manifolds, its \textbf{pull-back map} on 0-forms, $f^*: \Omega^0(\mathcal S) \to \Omega^0(\mathcal N)$ is defined by $(f^*(g))(p) = g(f(p))$ for any $g\in\Omega^0(\mathcal S)$ and $p\in \mathcal N$.  More concisely we may write $f^*\circ g = g\circ f$.

The \textbf{push-forward} map is then $f_*:\mathcal X(\mathcal N) \rightarrow \mathcal X(\mathcal S)$ defined by $(f_*(v))(g) = v(f^*(g))$ for any $v\in \mathcal X(\mathcal N)$. The push-forward generalises the notion of differential of a map. 
There is also a \textbf{pull-back} induced on the cotangent bundle by the map $f$, $f^{\ast}:\Omega^1(\mathcal S )\rightarrow \Omega^1(\mathcal N)$ is $$\big (f^{\ast}(\alpha) \big )v\; := \; \alpha\big (f_{\ast}v\big), $$
for $\alpha \in \Omega^1(\mathcal S)$. 

Hence the push-forward map acts on vector field and the pull-back on differential 1-forms. If $f:\mathcal M \rightarrow \mathcal M$ is a diffeomorphism, we define the \textbf{induced map} $\tilde f$ which maps tensor fields to tensor fields by: 
\begin{enumerate}
\item If $h \in C^{\infty}(\mathcal M)$, $\tilde f(h) := f^*(h) = h\circ f$.
\item If $v \in \mathcal X(\mathcal M)$, then $\tilde f(v) := f_*v = v\circ f^*$,
\item If $\omega \in \Omega^1(\mathcal M)$, then $\tilde f(\omega) =f^*(\omega) = \omega\circ f_*$.
\item If $T$, $S$ are tensor fields, then $\tilde f(T\otimes S):=\tilde f(T) \otimes \tilde f(S)$.
\end{enumerate}

The Lie derivative of a tensor field $T$ along a vector field $v$ is defined as 
$$\mathcal L_vT \; := \; \lim_{t\rightarrow 0}\frac{T-\tilde \sigma_t T}{t},$$
 where $\sigma$ is the local flow of $v$. The rate of change of any tensor,  along the trajectory of a Hamiltonian system with Hamiltonian $H$, may be shown to be equal to its Lie derivative in the direction of the flow: 
$$\frac{dT}{dt} \; = \; \mathcal L_{\hat H} T.$$ 
The formal solution of this equation is $T(t)=\exp \big ( t \mathcal L_{\hat H}\big)T(0)$, where 
$$\exp \big(t \mathcal L_{\hat H}\big ) 
\; := \; 
\sum_{k=0}^{\infty} \frac{1}{k!}t^k\big (\mathcal L_{\hat H} \big )^k.$$ 
Given a Hamiltonian vector field $\hat A$, the Lie derivative  defines the Hamiltonian trajectories followed by the physical system in phase space. More precisely, given a phase $p\in \mathcal M$, its evolution is given by the action of the the integral curve $\gamma : t \mapsto \exp(t \hat A)$ on $p$, which means that $\gamma(t)(p) \in \mathcal M$ gives the phase at time $t$ under the Hamiltonian system $(\mathcal{M},\omega, \hat A)$. 

\subsubsection{Maurer-Cartan Forms and Left-invariant Vector Fields}

A \textbf{Lie group} $G$ is a manifold which is also a group and on which the group operations are smooth. On a Lie group $G$, we define the \textbf{left/right translation}, $L_g,R_g :G \rightarrow G$, by $L_g(h)=gh$, $R_g(h)=hg$ for any $g \in G$. We say a vector field $v$ is \textbf{left-invariant} if $(L_g)_{\ast}v=v$ or in other words if for any $h \in G$, $(L_g)_{\ast}v_h=v_{gh}$. 
Such vector fields are constant, not in the sense of having constant components in a chart, but in the sense of being invariant under the maps induced by group multiplication. 
Any vector $\xi \in T_1G$ (here $1$ is the group identity) defines a unique left-invariant vector field $v$ by $v^{\xi}_g := (L_g)_{\ast}\xi$, which identifies $T_1G$ with the vector space of left-invariant vector fields on $G$, denoted $\mathfrak g$. 
 As there are $d$ linearly independent vectors in $T_1G$ there must be exactly $d$ linearly independent left-invariant vector fields $e_i$ for $i=1,\ldots,d$.
Moreover the commutator $[\cdot,\cdot ]$ of two left-invariant vector fields turns out to be a left-invariant vector field, it follows that it defines an operation (Lie Bracket) on $\mathfrak g$ which turns $\mathfrak g$ into a Lie algebra, called the \textbf{Lie algebra of $G$}. 

We define the \textbf{structure constants} $c^i_{jk}$ by $$[e_j,e_k] \; = \; c^i_{jk} e_i.$$

The 1-form fields $\theta^i$ dual to the left-invariant vector fields, $\theta^i(e_j) = \delta_{ij}$, are also left-invariant in the sense that $(L_g)^{\ast}\theta^i=\theta^i$. 
They are called Maurer--Cartan forms, and they satisfy the Maurer--Cartan relations $d\theta^i = -1/2 c^i_{jk} \theta^j\wedge\theta^k$.  The exterior product thus defined leads to the Chevalley--Eilenberg cohomolgy. \\

 It may be shown that the exponential map $\exp: \mathfrak g \rightarrow G$ is a local diffeomorphism on a neighbourhood $O\subset G$ of the identity $e$, with local inverse $\log$. This map can be used to define an operation on elements of $G$ near $e$ that are of the form $\exp( a)$ which outputs an element of $\mathfrak g$:  If $g_i =\exp(a_i)$, 
$$ g_1 \circ g_2 
\; := \;
 \log \big ( \exp(A_1) \exp(A_2) \big ) \in \mathfrak g.$$ 
The \textbf{Baker-Campbell-Hausdorff} (BCH) formula is a formal expansion for $g_1 \circ g_2$,  
$$g_1 \circ g_2 
\; = \;
 (A_1+A_2)+\frac{1}{2}[A_1,A_2]+\frac{1}{12}\Big( [A_1,[A_1,A_2]]-[A_2,[A_1,A_2]]\Big)+\cdots +$$
In particular we note that the Lie product $[\cdot,\cdot]$ of $\mathfrak g$ determines the group structure since from above we see that $g_1g_2=\exp(A_1)\exp(A_2)=\exp(A_3)$ where $A_3:=g_1 \circ g_2$. 

\subsection{Symplectic Integrators and Shadow Hamiltonians}\label{appendix:SHMC}

Given two Hamiltonian vector fields $\hat A,\hat B$ we can construct a curve $\gamma$ that is alternatively tangential to each vector field by composing their exponential maps (see Fig. \ref{fig:simulating_Hamiltonian}-b., \ref{fig:simulating_Hamiltonian}-c.): 
$$ \gamma(t) 
\; = \;
\Big( \exp(t \hat A/n)\exp (t \hat B/n) \Big)^n,$$ for some natural number $n$. This curve is called a symplectic integrator because it preserves the symplectic structure, that is $\omega =\gamma^{\ast}\omega$. A fundamental property of symplectic integrators is that from the BCH formula it can be shown this curve $\gamma$ is the integral curve of a Hamiltonian vector field $\hat D_{\epsilon}$: 
$$\gamma(t)
\; = \;
\Big( \exp(t \hat A/n)\exp (t \hat B/n) \Big)^n
\; = \;
\exp \big (t \hat D_{t/n}\big),$$ 
provided we choose $n$  large enough. The Hamiltonian $D_{\epsilon}$ corresponding to the Hamiltonian vector field $\hat D_{\epsilon}$ is called the \textbf{shadow Hamiltonian}.

We say a symplectic integrator is reversible if $\gamma(t) \gamma(-t) = e$. For example above this implies
$$ \exp(-t \hat A/n)\exp (-t \hat B/n) \exp(t \hat A/n)\exp (t \hat B/n) \; = \; e,$$
which means the symplectic steps $ \exp(t \hat A/n), \exp(t \hat B/n)$ commute. This is the case if and only if $[\hat A, \hat B]=0$. However we can easily construct non-trivial reversible symplectic integrators using symmetric symplectic integrators such as $\exp(t\hat A/2n)\exp(t \hat B/n) \exp (t\hat A/2n)$.

\subsection{Practical Integrators}
\label{appendix:practical_integrators}

It is usually not possible to find closed-form expressions for the integral curve of a Hamiltonian vector field. However there are some important special cases in which such an expression exists, and we consider an example here.
Darboux theorem implies it is possible to find a coordinate patch around any point that induces coordinates $(\bm q, \bm p)$ over which the symplectic structure takes the standard form $\omega =dq^i \wedge dp_i$. If $\hat A$ is a Hamiltonian vector field corresponding to the Hamiltonian $A$, in a Darboux patch it holds that 
$$\hat A \; = \; \frac{\partial A}{\partial p_i}\partial_{q^i}-\frac{\partial A}{\partial q^i}\partial_{p_i}.$$ 
If $\bm z(t)=\big(\bm q(t),\bm p(t)\big)$ is the integral curve of $\hat A$, we must have $\dot{\bm z}(t))=\hat A(\bm z(t))$, i.e., 
$$ \dot{ \bm q}(t)
\; = \;
\frac{\partial A}{\partial \bm p}(\bm z(t)), 
\quad \quad
\dot{ \bm p}(t)
\; = \; -\frac{\partial A}{\partial \bm q}(\bm z(t)).$$

When $A(\bm q,\bm p) = K(\bm p)$ or $A(\bm q,\bm p)=U(\bm q)$ we can find the exact solution. When $H(\bm q,\bm p)=K(\bm p)+U(\bm q)$ we can thus approximate the integral curves of $H$, by finding the integral curves of $K$ and $U$ separately and then using a symplectic integrator. Moreover from above we see these approximated integral curves will be the exact integral curves of a shadow Hamiltonian $H_{\epsilon}$ (and thus conserve its energy) which differs from $H$ by terms of order $\mathcal O(\epsilon)$ and $\mathcal O(\epsilon^2)$ if the integrator is reversible.

\subsection{Hamiltonian Mechanics on Lie Groups}
\label{appendix:Hamiltonia_mechanics_Lie_groups}

In this last section, we now define Hamiltonian mechanics on Lie groups. In order to do this, we will provide a natural definition of a kinetic energy and symplectic structure.

\subsubsection{Left-invariant Metric and Geodesics} 

We define the adjoint action at $\xi \in \mathfrak g$ as the map $ad_{\xi}:\mathfrak g \rightarrow \mathfrak g$ with $ ad_{\xi}(\nu):= [ \xi,\nu ]$.	The adjoint action satisfies $ad_{[\xi_1,\xi_2]}(\nu)=([ad_{\xi_1},ad_{\xi_2}])(\nu):=\big ( ad_{\xi_1} \circ ad_{\xi_2}-ad_{\xi_2}\circ ad_{\xi_1} \big )(\nu)$.

We say a Riemannian metric $\langle \cdot,\cdot \rangle$ on $G$ is left-invariant if 
$$\langle v_g, u_g \rangle _g  \; = \; \langle (L_h)_{\ast}v_g,(L_h)_{\ast}u_g \rangle_{hg}.$$
Thus to define a left-inavariant metric it is sufficient to specify a metric at the identity. The Cartan-Killing form on $\mathfrak g$ is defined as 
$$\langle \xi_1, \xi_2 \rangle 	 \; := \; \mathop{\rm Tr}( ad_{\xi_1} ad_{\xi_2}).$$
When it is non-degenerate it defines a left-invariant Riemannian metric and a kinetic energy $K=\frac{1}{2}\langle \cdot,\cdot \rangle$. We say a curve $g(t)$ is a geodesic if it is an extremal of the Lagrangian $$ \int K(\dot g) dt.$$

\subsubsection{Symplectic 2-form}  Any point in the cotangent space $T^{\ast}_gG$ may be written as $(g, p)$ where $p$ is the momentum. As $p$ is a 1-form we can expand it in terms of dual basis, $p=p_i\theta^i$. Hence a natural symplectic structure on $T^{\ast}G$ which respects the symmetries associated to base space $G$ is $\omega \; := \; -d(p_i\theta^i)$, which can also be written as $$\omega \; = \; \theta^i \wedge dp_i +\frac{1}{2}p_ic^i_{jk}\theta^j\wedge \theta^k.$$

For HMC we may then take the Hamiltonian to be of the form $H=K+U$, where $K$ is defined by the Cartan-Killing metric as shown above.


\subsection{Lagrangian Mechanics}
\label{appendix:Lagrangian_mechanics}

For a Newtonian particle with mass $m$ moving with velocity $v$ in a force field described by a potential energy $U$, Newton's equation states that $$m\frac{d^2 x^i}{dt^2}=-\frac{\partial U}{\partial x^i}.$$ Note
$$m\frac{dv^i}{dt}\; = \; -\frac{\partial U}{\partial x^i} 
 \quad  \Leftrightarrow \quad  
 \frac{d}{dt}\Big{(}\frac{\partial K(v)}{\partial v^i}\Big{)}+\frac{\partial U(x)}{\partial x^i} \; = \; 0
 \quad \Leftrightarrow  \quad 
 \frac{d}{dt}\Big{(}\frac{\partial \mathcal{L}}{\partial v^i}\Big{)}-\frac{\partial \mathcal{L}}{\partial x^i} \; = \; 0 $$ 
where the last equations are the Lagrange equations \big(a path $(x(t),v(t))$, $t\in(a,b)$ satisfies Euler-Lagrange equations if and only if it extremises the action $\int_a^b\mathcal{L}(x(t),v(t))dt$ \big). This shows that Newtonian physics can be reformulated as a Lagrangian system with Lagrangian $\mathcal{L}(x,v)=T(v)-U(x)$. 

Now consider an arbitrary manifold $\mathcal{M}$. The set of all tangent vectors, together with a map $\pi:T\mathcal M\rightarrow \mathcal M$ telling us which point in the manifold a vector in the bundle is ``over'', is called the tangent bundle $T \mathcal{M}:=\bigcup_{p\in \mathcal{M}} T_p \mathcal{M}$. Local coordinates $\bm x_V$ on $\mathcal{M}$ define local coordinates on then tangent spaces over $V$, $TV:=\bigcup_{p\in V} T_p \mathcal{M}$. Any vector $u\in TV$ may be written as $u=v^i\partial_i|_p$ which define a local patch by 
$$\bar{\bm x}_V(u) \; := \; (x^1(p),\ldots,x^d(p),v^1,\ldots,v^d),$$
which shows $T \mathcal{M}$ is a $2d$-dimensional manifold.

It is straightforward to generalise the above physical system and turn in into a coordinate-independent geometric theory by observing that the Lagrangian function is really an $\mathbb{R}$-valued function on the tangent bundle, and replacing the Euclidean inner product by an appropriate inner product: a \textbf{Lagrangian mechanical system} is a Riemannian manifold $(\mathcal M,g)$ together with a Lagrangian $\mathcal{L}:T\mathcal M\rightarrow \mathbb{R}$, where $\mathcal{L}=K-U$ and $K(u,u):=\frac{1}{2}g(u,u)$ and $U:\mathcal M\rightarrow \mathbb{R}$. 

Hamiltonian mechanics may also be viewed as a Legendre transformation of Lagrangian mechanics, which transforms the Lagrangian into a function on the cotangent bundle, the Hamiltonian.

\end{document}